\documentclass[lettersize,journal]{IEEEtran}
\IEEEoverridecommandlockouts
\usepackage{cite}
\usepackage{amsmath,amssymb,amsfonts}
\usepackage{graphicx}
\usepackage{textcomp}
\usepackage{colortbl}
\usepackage{xcolor}
\usepackage{cite}
\usepackage{amsfonts}

\usepackage{amsthm}
\usepackage{algorithm}
\usepackage{algpseudocode}

\usepackage{makecell}
\usepackage{mathtools}
\usepackage{booktabs}
\usepackage{multirow}
\usepackage{subfigure}
\usepackage{diagbox}
\usepackage{hyperref}
\newtheorem{Lem}{Lemma}
\newtheorem{remark}{Remark}

\usepackage{tikz}
\usetikzlibrary{positioning,decorations.pathreplacing}

\tikzset{
  box/.style={rectangle,draw,rounded corners,
              minimum width=1.2cm,minimum height=0.6cm,
              align=center,font=\scriptsize},
  cpu/.style={box,fill=blue!15},
  rc/.style={box,fill=green!15},
  switch/.style={box,fill=green!10},
  gpu/.style={box,fill=gray!10},
  server/.style={box,fill=gray!10},
  link/.style={-latex,thick},
}

\def \cz #1{\textcolor{black}{#1}}
\def \yp #1{\textcolor{black}{#1}}

\definecolor{blue}{RGB}{0, 0, 255}
\definecolor{black}{RGB}{0, 0, 0}
\definecolor{purple}{RGB}{128, 0, 128}

\def\BibTeX{{\rm B\kern-.05em{\sc i\kern-.025em b}\kern-.08em
    T\kern-.1667em\lower.7ex\hbox{E}\kern-.125emX}}
\begin{document}

% \title{Bandwidth-Aware Parameter Synchronization Topology Optimization for Decentralized Learning*\\
\title{Bandwidth-Aware Network Topology Optimization for Decentralized Learning
% \\
% {\footnotesize \textsuperscript{*}Note: Sub-titles are not captured in Xplore and
% should not be used}
% \thanks{Identify applicable funding agency here. If none, delete this.}
}

% \author{Anonymous Author(s)
% }

\author{%
    Yipeng~Shen, 
    Zehan~Zhu, 
    Yan~Huang, 
    Changzhi~Yan, 
    Cheng~Zhuo,
    and Jinming~Xu$^*$%
\thanks{
Y. Shen, Z. Zhu, C. Yan, and J. Xu are with the College of Control Science and Engineering, 
Zhejiang University, Hangzhou, China. C. Zhuo is with the College of Information Science and Electronic Engineering, Zhejiang University, Hangzhou, China.
Y. Huang is with the KTH Royal Institute of Technology, Stockholm, Sweden.
}
\thanks{
$^*$Corresponding author: Jinming Xu (e-mail: jimmyxu@zju.edu.cn).
}}

\maketitle

\begin{abstract}
% In recent years, decentralized learning has been extensively studied to cope with increasing dataset sizes and the complexity of neural network models. However, in order to achieve the consensus of model parameters, it is necessary to communicate model parameters or gradients frequently between computing nodes during the training process, which results in high communication overheads. 
% Many existing studies focus on designing a topology that can realize nearly-exact averaging, to speed up the convergence rate and improve the communication efficiency. However, these methods do not consider real-world bandwidth scenarios, and thus cannot effectively balance the per-iteration communication overhead and convergence rate. 

% Decentralized learning has become an enabling technique for accelerating parallel training. 
Network topology is critical for efficient parameter synchronization in distributed learning over networks. However, most existing studies do not account for bandwidth limitations in network topology design. In this paper, we propose a bandwidth-aware network topology optimization framework to maximize consensus speed under edge cardinality constraints. For heterogeneous bandwidth scenarios, we introduce a maximum bandwidth allocation strategy for the edges to ensure efficient communication among nodes. By reformulating the problem into an equivalent Mixed-Integer SDP problem, we leverage a computationally efficient ADMM-based method to obtain topologies that yield the maximum consensus speed. Within the ADMM substep, we adopt the conjugate gradient method to efficiently solve large-scale linear equations to achieve better scalability. Experimental results demonstrate that the resulting network topologies outperform the benchmark topologies in terms of consensus speed, and reduce the training time required for decentralized learning tasks on real-world datasets to achieve the target test accuracy, exhibiting speedups of more than $1.11\times$ and $1.21\times$
for homogeneous and heterogeneous bandwidth settings, respectively.
% exhibiting improvements of over 9.8\% and 17.9\% 

% In this paper, we formulate the design of network topology as an optimization problem that aims to maximize the consensus speed while subjecting to bandwidth-aware cardinality constraints on the edge set. Moreover, when it comes to heterogeneous bandwidth, we propose an algorithm that can allocate the possible maximum bandwidth for each edge, enabling fast communication per iteration. Furthermore, we solve the above problem for both scenarios of homogeneous and heterogeneous bandwidth leveraging the ADMM-based framework, which allows us to obtain the topology that yields fast consensus. Experimental results show that the obtained network topologies outperform the existing ones in terms of consensus speed, and reduce the time required for decentralized SGD (DSGD) to achieve the target test accuracy, exhibiting improvements of over 15.9\% and 23.5\% for homogeneous and heterogeneous bandwidth settings, respectively, when compared to existing network topologies.

% Experimental results show that the obtained network topologies outperform the existing ones in terms of consensus speed, and reduce the time required for decentralized SGD (DSGD) to reach the target test accuracy by over 15.9\% and 23.5\% for homogeneous and heterogeneous bandwidths, respectively, compared to the existing network topologies.

% and enable accelerated distributed training process for deep learning tasks to achieve the targeted test accuracy (\jm{**to be more specific}).

\end{abstract}


\section{Introduction}
% First, introduce the rapid development of distributed learning, and then the role of topology in distributed learning.
\IEEEPARstart{d}{eep} learning~\cite{lecun2015deep} has been widely used in various fields, including computer vision~\cite{jarvis1983perspective}, natural language processing~\cite{hirschberg2015advances}, speech recognition~\cite{han2017ese} and so on. As the complexity of tasks increases, the scale of datasets~\cite{deng2009imagenet} and the complexity of neural network models~\cite{fedus2022switch, zhang2022opt} used in deep learning are also increasing. To meet the training requirements of large-scale deep learning tasks, distributed learning has emerged~\cite{ben2019demystifying, tang2020communication}. It involves the collaborative efforts of multiple devices or computing nodes to achieve parallel processing of large datasets and complex models. However, in centralized training architectures such as Parameter Server (PS)~\cite{li2014communication} and All-Reduce~\cite{sergeev2018horovod}, frequent exchange of model parameters among nodes is needed to ensure the consensus of the model of nodes, leading to communication bottlenecks.

In order to reduce communication overhead, training algorithms based on decentralized architectures have been widely studied~\cite{koloskova2020unified, lian2017can, lu2021optimal, mcmahan2017communication}, where each computing node synchronizes parameters only with its neighboring nodes in the network topology (also referred to as parameter synchronization topology), known as partial averaging. These decentralized learning algorithms significantly reduce communication time per iteration compared to training based on centralized architectures \cite{lian2017can}. 
However, different from centralized counterparts, since partial averaging cannot guarantee instantaneous consensus of the model of nodes,  more iterations is usually required to reach the linear speedup stage~\cite{ying2021exponential}, which slows down the convergence speed.
% Thus, it is crucial to leverage the low communication overhead of decentralized learning while simultaneously accelerating its consensus speed.
In decentralized learning with limited bandwith, 
% the network topology determines the communication time per iteration and the consensus speed of the model. Specifically, 
the connectivity of the network topology determines the communication time per iteration, which, indeed, tends to have an inverse trend with consensus speed~\cite{ying2021exponential}.
% in homogeneous bandwidth scenarios \jm{where each node has the same bandwidth}, 
% while its weight matrix and connectivity influence the consensus speed. 
% the impact of network topology on the communication time per iteration tends to have an inverse trend with consensus speed
For instance, in sparse topologies, where nodes have lower connectivity, the communication time per iteration is short but the consensus speed is slow, and vice versa. Therefore, designing a network topology that can effectively balance the communication overhead per iteration and consensus speed is essential for improving the efficiency of distributed training. 

Existing network topologies are usually designed in an intuitive manner~\cite{ying2021exponential,nedic2018network,trevisan2017lecture,song2022communication,nachmias2008critical,benjamini2014mixing}, and they thus cannot effectively balance the communication overhead per iteration and consensus speed. For instance, the ring topology in~\cite{nedic2018network} has low communication overhead per iteration, but its consensus speed rapidly decreases as the number of nodes increases. In contrast, the exponential topology in~\cite{ying2021exponential} has a fast consensus speed, but as the number of nodes increases, the degree of nodes also increases, resulting in larger communication overhead per iteration. Moreover, the weights of edges in the network topologies based on intuitive design are typically assigned according to the degree of the nodes~\cite{nedic2018network}, which cannot guarantee the optimality of the weight matrix. There are also network topologies obtained through optimization-based methods, such as those in~\cite{xiao2004fast,sun2018weighted}. However, in \cite{xiao2004fast}, additional constraints are introduced on edge weights to simplify the solution process, thereby limiting the solutions to a subset of the solution space. Likewise, reference \cite{sun2018weighted} and most of the existing works design topologies for the scenario where the bandwidth of the nodes are homogeneous, which restricts their potential applications to more commonly encountered heterogeneous bandwidth scenarios.

% \yp{Additionally, existing work are mostly designed based on homogeneous network environment, which faces great performance degradation in actual heterogeneous scenarios.}

In this paper, we propose a bandwidth-aware network topology optimization framework for decentralized learning under the scenarios of homogeneous and heterogeneous bandwidth. The contributions of this work are summarized as follows:
 % To solve this problem, we utilize an ADMM-based solution framework (termed BA-Topo) to obtain the topologies that achieve fast consensus.
% In the optimization problems, the consensus rate (quantitative representation of the consensus speed) is used as the optimization objective and the bandwidth-aware cardinality constraints on the edge set are used to achieve the balance between the per-iteration communication time and the consensus speed. Specifically, for the heterogeneous bandwidth scenario, we design a bandwidth allocation algorithm to allocate the possible maximum bandwidth for each edge. Furthermore, to solve the network topology optimization problems, we introduce auxiliary variables to transform the optimization problems and utilize an ADMM-based solution framework, which allows us to obtain the topologies (termed BA-Topo) that achieve fast consensus. \hy{**This paragraph too detailed and overlaps with the following bullets**}
% In summary, the contributions of this work are as follows:

\begin{itemize}
% optimization problem, solver, experiments
    \item We formulate the design of network topology as a constrained optimization problem that aims to maximize the consensus speed. In particular, we explicitly introduce bandwidth-aware cardinality constraints on the edge set for both homogeneous and heterogeneous scenarios. 
    % For heterogeneous bandwidth settings, we further incorporate constraints arising from practical physical systems, yielding a practical formulation that faithfully reflects real-world architectural limitations. Moreover, we propose a maximum-bandwidth allocation algorithm for each edge under such constraints, enabling fast communication among nodes.
    For heterogeneous settings, the bandwidth constraints imposed by real systems are included to reflect hardware-level limitations. Besides, we propose an allocation algorithm that determines the maximum number of edges per node, enabling faster communication among nodes.

    % \item We formulate the design of network topology as a constrained optimization problem that aims to maximize the consensus speed. In particular, we explicitly introduce bandwidth-aware cardinality constraints on the edge set, making it applicable to both homogeneous and heterogeneous scenarios. Moreover, for scenarios with heterogeneous bandwidth, we propose a maximum bandwidth allocation algorithm for each edge, resulting in fast communication among nodes. 
    % Solving the optimization problem allows us to effectively balance communication time per iteration and consensus speed. 
    \item We reformulate the network topology design problem to an equivalent Mixed-Integer SDP problem that can be solved under an ADMM-based framework.
    % first introduce auxiliary variables to decouple the cardinality constraints from the linear matrix inequality (LMI) constraints,
    % thereby improving its solvability. 
    Leveraging a properly designed ADMM-based optimization algorithm with conjugate gradient, we obtain topology from the full solution space that yields faster consensus and better scalability than existing works for both scenarios of homogeneous and heterogeneous bandwidth.
    % \cz{Compared with the existing optimization-based methods, we obtain network topologies from the full solution space, rather than in the subset of the solution space.}
    % auxiliary variables to achieve equivalent transformations of the optimization problems, and then use the ADMM-based framework to solve the above network topology optimization problems, optimizing both the topology structure and the weight matrix simultaneously. 
    \item We conduct comprehensive experiments in both homogeneous and heterogeneous scenarios to validate the consensus speed of the obtained topology (termed BA-Topo) and its effectiveness in decentralized learning. The experimental results show that BA-Topo outperforms the existing topologies in terms of consensus speed, and reduces the time required for decentralized SGD (DSGD) to achieve a target test accuracy, exhibiting speedups of more than $1.11\times$ and $1.21\times$ for real homogeneous and heterogeneous bandwidth settings, respectively.
\end{itemize}

The rest of the paper is organized as follows. Section~\ref{section2} reviews related works. Section~\ref{section3} introduces relevant background knowledge. Section~\ref{section_four} describes the network topology optimization problems. Section~\ref{section5} outlines the framework for solving these optimization problems. Section~\ref{section6} presents the experimental results. Finally, Section~\ref{section7} concludes the paper and provides an outlook for future research.

\section{Related Works} \label{section2}
% Introduce related work, including distributed learning, the impact of topology, and existing topology design methods(direct design and optimization-based methods)
Communication in decentralized learning typically involves two types of topologies: the underlying physical connection topology of GPUs and the logical topology for parameter synchronization. The underlying physical topology determines the connectivity among GPU servers, that is, the switch fabric used for communication between any two servers. Commonly used
physical topologies include FatTree~\cite{al2008scalable} and BCube~\cite{guo2009bcube}. The parameter synchronization topology which is the focus of this work, instead, determines which GPUs or servers can communicate with each other on top of the underlying physical topology. Current research on parameter synchronization topology can be categorized into two amin streams: intuition-based design and optimization-based design.

\textbf{Intuition-based design}. Angelia et al.\cite{nedic2018network} conduct theoretical research on the impact of various topologies (such as ring, grid, torus, etc.) on distributed training, demonstrating that these topologies maintain a constant degree regardless of the number of nodes. However, as the number of nodes increases, the convergence speed of these topologies decreases rapidly, leading to inefficient parameter synchronization. The hypercube topology proposed by Trevisan et al.\cite{trevisan2017lecture} achieves a good balance between the degree of nodes and consensus speed, with a logarithmic relationship between consensus speed and the number of nodes. However, this topology requires the number of nodes to be a power of 2, making it impractical when the number of nodes does not meet this condition. As an extension of the hypercube topology, the exponential topology~\cite{ying2021exponential} ensures both sparse communication and faster convergence speed. In contrast, Song et al.\cite{song2022communication} design directed and undirected topologies based on a series of some basic topologies, achieving a constant degree and convergence speed independent of the number of nodes. Moreover, Nachmias et al.\cite{nachmias2008critical} and Benjamini et al.\cite{benjamini2014mixing} investigate random topologies where each edge is randomly activated with a certain probability. However, these methods may generate relatively dense topologies, leading to highly uneven degrees among nodes. Additionally, there is research on parameter synchronization topologies based on the All-Reduce architecture\cite{sergeev2018horovod,jia2018highly,tanakaimagenet,iandola2016firecaffe,zhao2013butterfly}, which is, however, not within the scope of this paper. From the above, it is evident that intuition-based design of parameter synchronization topology faces challenges in obtaining the optimal weight matrix, making it difficult to improve the consensus speed of the model. Furthermore, existing literature often assumes a homogeneous network environment, leading to performance degradation in actual heterogeneous scenarios.

\textbf{Optimization-based design.}  In the context of this paper, there is a relatively limited body of research on optimization-based methods. Xiao et al.\cite{xiao2004fast} propose an optimization problem of the weight matrix to maximize the average consensus speed on undirected topologies, and they reformulate the problem as an SDP problem which can be solved efficiently. To simplify the solution process, they set the weight of each edge in the topology to be the same, resulting in a weight matrix that is only a solution within a subset of the solution space. Sun et al.\cite{sun2018weighted} address the consensus problem in multi-agent systems and describe the network topology optimization problem with constraints on the number of edges. They propose a customized method based on the ADMM technique~\cite{boyd2011distributed} to solve this optimization problem, achieving the simultaneous design of both network topology structure and edge weights. However, since they focus on network topologies applied to continuous-time problems, their approach is not directly applicable to decentralized learning in our setting. There have been also few works that employ different optimization objectives. For instance, Marfoq et al.\cite{marfoq2020throughput} design an optimization problem for federated learning to maximize system throughput and propose a corresponding topology generation algorithm. Also, some research focuses on optimization of parameter synchronization topology in data heterogeneous scenarios\cite{dandi2022data,le2023refined}, aiming to improve the convergence speed of decentralized learning.

% \yp{There is also a lack of research on heterogeneous bandwidth scenarios}.
% \cz{In this paper, we take consensus speed as the optimization objective, so it can be seen as an improvement of \cite{xiao2004fast} and \cite{sun2018weighted}}.

% From the above,  Additionally, methods based on solving optimization problems often simplify the optimization objectives or processes, leading to a local optimum within a limited space. Moreover, these methods typically do not consider communication overhead in the design of optimization problems. Therefore, we propose the bandwidth-aware network topology optimization method in this work, achieving a better balance between per-iteration communication time and consensus speed.

\section{Preliminaries} \label{section3}
% Introduce the related symbols, the conditions that topology needs to meet, and the consistency problem studied in this paper

\textbf{Notations.} Let $\mathcal{G} \left( \mathcal{N} ,\mathcal{E} \right) $ denote a graph consisting of $n$ nodes and $e$ edges with $\mathcal{N}$ representing the node set and $\mathcal{E}$ representing the edge set, where each edge $\left\{ i,j \right\} \in \mathcal{E}$ is an undirected edge and the neighbors of node $i$ are denoted by $\mathcal{N} _i=\left\{ j|\left\{ i,j \right\} \in \mathcal{E} \right\}$. The weight matrix of $\mathcal{G}$ is denoted as $W$. The network topology studied in this paper includes both $\mathcal{G}$ and $W$. For a vector $\boldsymbol{x}$, $\mathbf{Card}\left( \boldsymbol{x} \right) $ represents the number of nonzero elements in the vector, and $\mathbf{Diag}\left( \boldsymbol{x} \right) $ represents a diagonal matrix whose diagonal components are the elements of $\boldsymbol{x}$. For a matrix $P$, $\mathbf{diag}\left( P \right) $ represents a vector whose components are the diagonal elements of the matrix $P$. For two matrices $X_1$ and $X_2$, the notation $X_1 \succcurlyeq  X_2$ represents that the eigenvalues of $X_1$ are all bigger than or equal to the corresponding eigenvalues of $X_2$, $X_1 \curlyeqprec X_2$ represents that the eigenvalues of $X_1$ are all smaller than or equal to the corresponding eigenvalues of $X_2$, and $X_1 \sim X_2$ indicates the similarity between $X_1$ and $X_2$. For two vectors $\boldsymbol{x}_1$ and $\boldsymbol{x}_2$, $\boldsymbol{x}_1 \odot \boldsymbol{x}_2$ represents the Hadamard product, where the elements of two vectors at corresponding positions are multiplied, with the result still being a vector, and $\left< \boldsymbol{x}_1,\boldsymbol{x}_2 \right> $ represents the inner product of vectors. 
For two sets $\mathcal{C}_1$ and $\mathcal{C}_2$, $\mathcal{C}_1 \otimes \mathcal{C}_2$ represents their cartesian product.
The notation $\lfloor \cdot \rfloor$ represents rounding down the elements, and $\left\| \cdot \right\| _F$ represents the Frobenius norm of a vector or matrix. $\mathrm{vec}$ represents the vectorization of a matrix by stacking its columns on one another from left to right, and $\mathrm{abs}$ means that takes absolute values for each element. $\mathrm{Proj}_{\varOmega}\left( \cdot \right) $ denotes the projection of the element to the feasible domain $\varOmega$. $\mathbf{0}_n$ denotes an n-dimensional vector with all zeros while $\mathbf{1}_n$ with all ones, and $\mathbf{0}_{n\times m}$ denotes a matrix of $n$ rows and $m$ columns with all zeros.

\textbf{Consensus rate as optimization objective.} In decentralized learning, consensus rate is widely used to quantify the speed of consensus of the model parameters. Let $\boldsymbol{x}_k =\left[ \boldsymbol{x}_{1,k} ,...,\boldsymbol{x}_{n,k} \right]^T$ with $\boldsymbol{x}_{i,k}\in \mathbb{R} ^q, i\in [n]$ denoting the model parameters at the $k$-th iteration. The parameter synchronization process can be represented as follows:
\begin{equation}
    \label{tab:parameter_synchronization_process}
    \begin{aligned}
        \boldsymbol{x}_{k+1} =W\boldsymbol{x}_k , \ W\in \mathcal{S},
    \end{aligned}
\end{equation}
where $\mathcal{S} =\left\{ W\in \mathbb{R} ^{n\times n}|W_{ij}=0 \  \mathrm{if} \left\{ i,j \right\} \notin \mathcal{E} \ \mathrm{and} \  i\ne j \right\}$, and $W$ needs to satisfy the following conditions:
\begin{equation*}
    \mathbf{1}_{n}^{T}W=\mathbf{1}_{n}^{T}, \,\,W\mathbf{1}_n=\mathbf{1}_n, \,\, \rho \left( W-\frac{\mathbf{1}_n\mathbf{1}_{n}^{T}}{n} \right) <1,
\end{equation*}
% \begin{equation}
%     \label{tab:column_random}
%     W\mathbf{1}_n=\mathbf{1}_n,
% \end{equation}
% \begin{equation}
%     \label{tab:spectral_radius}
%     \rho \left( W-\frac{\mathbf{1}_n\mathbf{1}_{n}^{T}}{n} \right) <1,
% \end{equation}
where $\rho$ represents the spectral radius of the matrix. Then, the consensus rate can be denoted by the asymptotic convergence factor~\cite{xiao2004fast}:
\begin{equation}
    \label{tab:asymptotic_convergence_factor}
    r_{\mathrm{asym}}(W)=\mathop {\mathrm{sup}} \limits_{\boldsymbol{x}_0\ne \bar{\boldsymbol{x}}}\lim_{k\rightarrow \infty} \left( \frac{\parallel \boldsymbol{x}_k-\bar{\boldsymbol{x}}\parallel _2}{\parallel \boldsymbol{x}_0-\bar{\boldsymbol{x}}\parallel _2} \right) ^{1/k},
\end{equation}
where $\bar{\boldsymbol{x}}=\frac{\mathbf{1}}{n}\sum_{i=1}^n{\boldsymbol{x}_{i,0}}$.

According to the characteristics of the weight matrix, the consensus rate is related to the eigenvalue of the matrix, expressed as follows:
\begin{equation}
\label{tab:asymptotic_convergence_factor_newest}
    r_{\mathrm{asym}}(W)=\max \left\{ \left| \lambda _2\left( W \right) \right|,\left| \lambda _n\left( W \right) \right| \right\}.
\end{equation}
The smaller the value of $r_{\mathrm{asym}}(W)$, the faster the consensus speed of the model will be.

\section{Problem Formulation}\label{section_four}
% Introduce the topology optimization problem of communication, how to consider the communication-related constraints, and the problem design in homogeneous and heterogeneous bandwidth scenarios.

% According to the introduction in Section~\ref{section_four}, the network topology optimization problems can be designed with consensus rate as the optimization objective, so that the optimized topology can accelerate the convergence of the model. In addition, in practical decentralized learning scenarios, topology design needs to consider communication overhead to achieve a balance between consensus rate and communication time per iteration. 
In this section, we construct the bandwidth-aware network topology optimization problems with consensus rate as the optimization objective for homogeneous bandwidth scenarios and heterogeneous bandwidth scenarios, respectively. 

\subsection{Homogeneous Bandwidth}

% \hy{(Maybe we should motivate this scenario in one sentence first. Why we need to consider homogeneous bandwidth? List some references.)}
\cz{The existing research usually evaluates the communication time per iteration based on the degree of nodes~\cite{ying2021exponential, takezawa2023beyond}, which is valid when the available bandwidth of each node is the same, and we refer to this scenario homogeneous bandwidth scenario.} Due to the symmetry of the undirected topology, the number of edges and the degree of nodes are highly correlated. Therefore, in order to ensure the sparsity of the topology, we design the following network topology optimization problem using the number of edges as a constraint:
\begin{equation}
    \label{tab:homogeneous-problem-1}
    \begin{matrix}
        \underset{\mathbf{g}}{\min}&		\max \left\{ \left| \lambda _2\left( W \right) \right|,\left| \lambda _n\left( W \right) \right| \right\}\\
        s.t.&		\begin{array}{c}
        W=W^T, \ W\mathbf{1}=\mathbf{1},\\
        \mathbf{g} \geqslant 0, \ \mathbf{Card}\left( \mathbf{g} \right) \leqslant r,\\
    \end{array}\\
    \end{matrix}
\end{equation}
where $\mathbf{g}\in \mathbb{R}^{|\mathcal{E}|}$ is the vector consisting of the weights of the edges in the topology, ${|\mathcal{E}|}=\frac{n\left( n-1 \right)}{2}$, $r\le {|\mathcal{E}|}\in \mathbb{N}$ is a constraint on the number of edges. Gao et al.\cite{gao2011cardinality} call optimization problems shaped like Eq.~\eqref{tab:homogeneous-problem-1} Cardinality-Constrained Optimization Problems (CCOPs), where $\mathbf{Card}\left( \mathbf{g} \right) \le r$ is the cardinality constraint.

In order to reduce the difficulty of solving the problem, we first transform Eq.~\eqref{tab:homogeneous-problem-1} based on the Laplacian matrix of the topology. According to the relationship between the Laplacian matrix and the weight matrix:
\begin{equation}
    \label{tab:laplacian-matrix-weight-matrix}
    W=I-L=I-A\mathbf{Diag}\left( \mathbf{g} \right) A^T,
\end{equation}
where $A\in \mathbb{R} ^{n\times m}$ is the incidence matrix defined as follows:
% , representing the relationship between nodes and edges and is defined as follows:
\begin{equation}
    \label{tab:incidence matrix}
    A_{il}=\left\{ \begin{matrix}
	1&		\mathrm{if} \ \mathrm{edge} \ l\,\,\mathrm{starts} \ \mathrm{from}\ \mathrm{node}\ i,\\
	\begin{array}{c}
	-1\\
	0\\
\end{array}&		\begin{array}{c}
	\mathrm{if}\ \mathrm{edge}\ l\,\,\mathrm{ends}\ \mathrm{at} \ \mathrm{node}\ i,\\
	\mathrm{otherwise}.\\
\end{array}\\
\end{matrix} \right. 
\end{equation}

For the undirected topology studied in this paper, it is sufficient to arbitrarily assign the direction of each edge, resulting in the same Laplace matrix.
% it is sufficient that the direction of each of its edges is arbitrarily given, yielding the same Laplace matrix. 
% From the above definition of the incidence matrix $A$,
Noticing that the sum of each column of $A$ equals zero, the matrix $W$ defined in Eq.~\eqref{tab:laplacian-matrix-weight-matrix} automatically becomes symmetric and doubly stochastic. And the eigenvalues of $L$ satisfy:
\begin{equation}
    \label{tab:eig-of-laplacian-matrix}
    0=\lambda _n\left( L \right) <\lambda _{n-1}\left( L \right) \leqslant ...\leqslant \lambda _1\left( L \right) <2.
\end{equation}
Furthermore, Eq.~\eqref{tab:homogeneous-problem-1} can be transformed to:
\begin{equation}
    \label{tab:homogeneous-problem-2}
    \begin{matrix}
        \underset{\mathbf{g}}{\min}&		\max \left\{ \left| 1-\lambda _{n-1}\left( L \right) \right|,\left| 1-\lambda _1\left( L \right) \right| \right\}\\
        s.t.&		\mathbf{g}\geqslant 0,\ \mathbf{Card}\left( \mathbf{g} \right) \leqslant r,\\
    \end{matrix}.
\end{equation}

Since the eigenvalues of the Laplace matrix in Eq.~\eqref{tab:homogeneous-problem-2} are not explicit functions of the optimization variable $\mathbf{g}$ and solving for the eigenvalues of the matrices with the optimization variables is very complicated, we transform the optimization problem based on Lemma~\ref{lemma_eig-laplacian}.
\begin{Lem}[Proposition 1 in \cite{dai2011optimal}]
\label{lemma_eig-laplacian}
For a Laplace matrix $L$ with eigenvalues satisfying Eq.~\eqref{tab:eig-of-laplacian-matrix}, if $\alpha \geqslant \lambda _{n-1}\left( L \right)$, then $L+\frac{\alpha \mathbf{1}\mathbf{1}^ T}{n}\succcurlyeq \lambda _{n-1}\left( L \right) I$.
\end{Lem}

With the help of Lemma \ref{lemma_eig-laplacian}, the problem of minimizing the eigenvalues can be transformed into addressing a Linear Matrix Inequality (LMI) by introducing auxiliary variables, which in turn transforms Eq.~\eqref{tab:homogeneous-problem-2} into an optimization problem of the following form:
\begin{equation}
    \label{tab:homogeneous-problem-3}
    \begin{matrix}
        &	\underset{\mathbf{g},\tilde{\lambda}}{\min} \,\,	-\tilde{\lambda}\\
        s.t.&		\begin{array}{c}
        \mathbf{g}\geqslant 0,\,\,\tilde{\lambda}>0,\,\,\mathbf{Card}\left( \mathbf{g} \right) \leqslant r,\\
        L+\frac{\alpha \mathbf{1}\mathbf{1}^T}{n}\succcurlyeq  \tilde{\lambda} I,\,\,
        L\curlyeqprec \left( 2-\tilde{\lambda} \right) I,\,\,
        \mathbf{diag}\left( L \right) \leqslant \mathbf{1}.
    \end{array}\\
    \end{matrix}
\end{equation}
These two LMI constraints on the Laplace matrix above ensure that $\lambda _{n-1}\left( L \right)$ and $\lambda _1\left( L \right)$ are simultaneously close to one, thus minimizing the asymptotic convergence factor, i.e. maximizing the consensus rate. The last inequality constraint ensures that all the elements of the resulting weight matrix are not negative, since the weights used for parameter synchronization in decentralized learning are not negative.

% \hy{Highlight the challenges of this problem here.}

\begin{remark}
We formulate the design of network topology with homogeneous bandwidth as an optimization problem (c.f., Eq.~\eqref{tab:homogeneous-problem-3}) that can be solved asymptotically. Moreover, the cardinality constraint on the number of edges is introduced to ensure the sparsity of the topology. The resulting network topology will be shown to be able to provide a better balance between the consensus rate and the communication time per iteration than existing network topologies.
\end{remark}

\subsection{Heterogeneous Bandwidth}
We consider a practical heterogeneous bandwidth scenario, where the available bandwidth varies across nodes, e.g., intra-server links, and inter-server switch ports in multi-GPU platforms.
This section addresses the topology design problem in three typical heterogeneous bandwidth settings: 
i) heterogeneous node bandwidth, 
ii) heterogeneous intra-server link bandwidth, and 
iii) heterogeneous inter-server switch-port bandwidth.
Leveraging optimal bandwidth allocation, our approach achieves an effective trade-off between consensus speed and the per-iteration communication time.

To encode heterogeneous bandwidth constraints in a unified framework, we introduce
an edge-capacity constraint vector $\boldsymbol{e}$ and an incidence matrix $M$
to reformulate the cardinality constraint in 
Eq.~\eqref{tab:homogeneous-problem-3}. 
By introducing binary variables $\mathbf{z}$, the cardinality constraint can be rewritten as $\mathbf{g}\leqslant \mathbf{z}, M\mathbf{z}=\boldsymbol{e}$, where $\mathbf{z}$ indicates which logical edges are selected, and
$\mathbf{g}$ contains the associated edge weights, while $\boldsymbol{e}\in\mathbb{N}^{q}$ specifies the edge-capacity limits of
$q$ physical constraints (e.g., node degrees, intra-server link capacities, or 
switch-port capacities). As a result, the optimization problem~\eqref{tab:homogeneous-problem-3} becomes:
\begin{equation}
    \label{tab:heterogeneous-problem-2}
    \begin{matrix}
        &		\underset{\mathbf{g},\tilde{\lambda}}{\min} \,\, -\tilde{\lambda}\\
        s.t.&		\begin{array}{c}
        \mathbf{g}\geqslant 0,\,\,\tilde{\lambda}>0,\\
        \mathbf{z}\in \left\{ 0,1 \right\} ^{|\mathcal{E}|},\,\,\mathbf{g}\leqslant \mathbf{z},\,\,M\mathbf{z}=\boldsymbol{e},\\
        L+\frac{\alpha \mathbf{1}\mathbf{1}^T}{n}\succcurlyeq  \tilde{\lambda} I,\,\,
        L\curlyeqprec \left( 2-\tilde{\lambda} \right)  I,\,\,
        \mathbf{diag}\left( L \right) \leqslant \mathbf{1}.
    \end{array}
    \end{matrix}
\end{equation}
Note that the incidence matrix $M\in\{0,1\}^{q\times {|\mathcal{E}|}}$
encodes \emph{the mapping from logical edges to physical bandwidth constraints}. In particular, each row
corresponds to a physical constraint and each column corresponds to a logical
edge, i.e.,
\begin{equation}
\label{tab:M_definition}
M=\begin{bmatrix}
m_{11} & m_{12} & \cdots & m_{1e}\\
m_{21} & m_{22} & \cdots & m_{2e}\\
\vdots & \vdots & \ddots & \vdots\\
m_{c1} & m_{c2} & \cdots & m_{ce}
\end{bmatrix}\in\{0,1\}^{q\times {|\mathcal{E}|}},
\end{equation}
where the $i$-th row vector $\boldsymbol{m}_i = (m_{i1},\ldots,m_{ie})^{T}$ indicates which logical edges consume the $i$-th physical resource, whose capacity limit is given by the $i$-th element $e_i$ of
$\boldsymbol{e}$. Thus, the constraint $M\mathbf{z}=\boldsymbol{e}$ ensures that the activation of logical edges adheres to all physical bandwidth limits.
% \begin{remark}
%     \jm{The incorporation of node-level edge-capacity constraints (c.f., $M$ and $\boldsymbol{e}$ in Eq.~\eqref{tab:heterogeneous-problem-2}) ensures maximum utilization of available bandwidth per edge, thereby achieving fast communication per iteration.
%     The resulting network topology attains both fast consensus speed and low communication time per iteration, which is not realizable for existing topologies.}
% \end{remark}

Now, we show that, by specifying $M$ and $\boldsymbol{e}$ appropriately, one can model heterogeneous bandwidth conditions at the node level, within servers, and across server switch ports, all within a unified topology optimization framework.

\subsubsection{Node-Level Bandwidth Heterogeneity}
We first consider heterogeneous bandwidth across nodes.
We propose an edge-capacity allocation algorithm, as shown in Algorithm~\ref{Algorithm-Bandwidth-Allocation}.
Given the available bandwidth of each node (or link or port; we use nodes for example), the algorithm determines the number of edges per node to maximize the bandwidth per edge under the given bandwidth constraints.
Suppose the bandwidths of $n$ nodes are given by $\boldsymbol{b}=\left( b_1,\ldots ,b_n \right)^T$, the total number of edges to be allocated is $r$, and the upper limit on the number of edges incident to each node is $\bar{\boldsymbol{e}}=\left( \bar{e}_1,\ldots ,\bar{e}_n \right)^T$.
\begin{center}
\begin{algorithm}[!htbp]
      \caption{Bandwidth-Aware Edge-Capacity Allocation}
      \label{Algorithm-Bandwidth-Allocation} 
        \begin{algorithmic}[1]
        \Require{$\boldsymbol{b}=\left( b_1,... ,b_n \right)^T,\, r,\, \bar{\boldsymbol{e}}=\left( \bar{e}_1,... ,\bar{e}_n \right)^T$.}
        \Ensure{Unit bandwidth $b_{unit}$ and number of edges connected to each node $\boldsymbol{e}=\left( e_1,...,e_n \right)^T$.}
        \State Initialize the unit bandwidth, the number of edges on each node, and the total number of edges:
        \begin{equation}
            \begin{aligned}
            \label{tab:initialize-edges}
            b_{unit}&=\underset{i\in \left\{ 1,...,n \right\}}{\min}\,\,b_i,
            \\
            e_i&=\min \left( \lfloor \tfrac{b_i}{b_{unit}} \rfloor ,\bar{e}_i \right) ,\ i=1,...,n,
            \\
            \mathrm{edge}\_\mathrm{count}&=\frac{1}{2}\sum_{i=1}^n{e_i}.
            \end{aligned}
        \end{equation}
        \While{$\mathrm{edge}\_\mathrm{count}<r$}
            \State Calculate the new unit bandwidth and update the number of edges on each node:
            \begin{equation}
                \begin{aligned}
                b_{unit}&=\underset{i\in \left\{ 1,...,n \right\}}{\max}\,\,\frac{b_i}{e_i+1},\\
                e_i&=\min \left( \lfloor \tfrac{b_i}{b_{unit}} \rfloor ,\bar{e}_i \right) ,\ i=1,...,n.
                \end{aligned}
            \end{equation}
            \State Calculate the total number of edges:
            \begin{equation}
                \label{tab:total-edge-number}
                \mathrm{edge}\_\mathrm{count}=\frac{1}{2}\sum_{i=1}^n{e_i}.
            \end{equation}
        \EndWhile
        \If{$\mathrm{edge}\_\mathrm{count}>r$}
            \State Keep subtracting the number of edges on the node with the highest number of edges by 1 and calculate the total number of edges using Eq.~\eqref{tab:total-edge-number} until the total number of edges is equal to $r$.
        \EndIf
        \State \textbf{return} $b_{unit},\ \left( e_1,...,e_n \right)^T$ 
        \end{algorithmic}
\end{algorithm}
\end{center}

The unit bandwidth $b_{unit}$, defined as the minimum bandwidth among all edges, determines the communication time per iteration.
Lines 6--8 ensure that the total number of edges meets the constraint $r$ while distributing edge-capacity across nodes.
Algorithm~\ref{Algorithm-Bandwidth-Allocation} maximizes the bandwidth assigned to each edge under the node-level bandwidth limits.

To embed these node-level capacities into the unified formulation~\eqref{tab:heterogeneous-problem-2}, we let the number of physical constraints be $q=n$ and define the edge-capacity constraint vector as
$\boldsymbol{e} = (e_1,\ldots,e_n)^{T}$, where $e_i$ is the number of logical edges allocated to node $i$ by Algorithm~\ref{Algorithm-Bandwidth-Allocation}.
For each node $i$, we construct an incidence vector $\boldsymbol{m}_i\in \mathbb{R}^{|\mathcal{E}|}$ to identify whether each logical edge is incident to node $i$:
\begin{equation}
    \label{tab:mask_i}
    m_{il}=\left\{ 
    \begin{array}{ll}
	1, & \mathrm{if\ edge}\ l\ \mathrm{is\ connected\ to\ node}\ i,\\[3pt]
	0, & \mathrm{otherwise},
    \end{array} \right. 
\end{equation}
where $m_{il}$ is the $l$-th element of $\boldsymbol{m}_i$, $l=1,\ldots,|\mathcal{E}|$.
In this case, $\boldsymbol{m}_i=\mathrm{abs}\left( A_{i,:} \right)$, where $A_{i,:}$ denotes the $i$-th row of the incidence matrix $A$ defined in Eq.~\eqref{tab:incidence matrix}.
By stacking all node-level masks, we obtain the incidence matrix
\begin{equation}
M
=
\begin{bmatrix}
\boldsymbol{m}_1^{T}\\
\vdots\\
\boldsymbol{m}_n^{T}
\end{bmatrix}
=\mathrm{abs}(A)
\in\{0,1\}^{n\times {|\mathcal{E}|}},
\end{equation}
whose $i$-th row enforces the edge-capacity constraint associated with node $i$.
Applying this $M$ and $\boldsymbol{e}$ in Eq.~\eqref{tab:heterogeneous-problem-2} yields the node-level bandwidth heterogeneity formulation.

\subsubsection{Intra-Server Link Bandwidth Heterogeneity}

We next consider heterogeneous bandwidth across intra-server physical links.
The intra-server interconnect is modeled as a hierarchical tree topology $\mathcal{T} = (\mathcal{V},\mathcal{L})$, where $\mathcal{V}$ denotes the set of internal switching or routing components and leaf computational devices, and $\mathcal{L}$ denotes the set of physical links connecting them.
Each physical link $\ell\in\mathcal{L}$ is associated with an available bandwidth $b_{\ell}>0$ and, due to hardware-level capacity constraints, can support only a limited number of logical edges.
We denote this per-link capacity by $e_{\ell}\in\mathbb{N}$.

In the unified framework~\eqref{tab:heterogeneous-problem-2}, the intra-server edge-capacity constraint vector $\boldsymbol{e}$ is defined as $\boldsymbol{e} = (e_{\ell})_{\ell\in\mathcal{L}}$,
where each entry specifies the maximum number of logical edges that can be assigned to a physical link.
The incidence matrix $M$ is defined as $M \in \{0,1\}^{|\mathcal{L}|\times |\mathcal{E}|}$,
where the entry in the row corresponding to link $\ell$ and the column corresponding to logical edge $k\in\mathcal{E}$ is defined as
\begin{equation}
M_{\ell k} =
\begin{cases}
1, & \text{if logical edge } k \text{ traverses link } \ell,\\[3pt]
0, & \text{otherwise}.
\end{cases}
\end{equation}
Thus, the row vector $\boldsymbol{m}_{\ell}$ encodes the set of logical edges that utilize physical link $\ell$.
With these definitions, the per-row constraint in $M\mathbf{z}=\boldsymbol{e}$ enforces per-link edge-capacity limits, and solving~\eqref{tab:heterogeneous-problem-2} with this $M$ and $\boldsymbol{e}$ yields the intra-server link Bandwidth heterogeneity formulation.

\subsubsection{Inter-Server Switch-Port Bandwidth Heterogeneity}

We finally consider heterogeneous bandwidth across inter-server switch ports.
We model the inter-server communication fabric as a multi-layer switching hierarchy in which servers 
are connected through several tiers of switches. Let the switch hierarchy consist 
of $k$ layers, indexed from $0$ (nearest to servers) to $k-1$ (highest level). 
Each switch at layer $i$ is equipped with multiple ports, and each port corresponds 
to a physical communication link with available bandwidth $b_{s_i}>0$. 
Due to hardware concurrency and bandwidth limitations, each port can support only 
a limited number of logical edges; we denote this per-layer port capacity by 
$e_{s_i}\in\mathbb{N}$.

Let $n_i$ denote the number of ports at layer $i$, and collect the per-layer 
port capacities into $\boldsymbol{e}_{s}=\left(e_{s_0},\,e_{s_1},\,\ldots,\,e_{s_{k-1}}\right)^{T}$.
In the unified formulation~\eqref{tab:heterogeneous-problem-2}, the 
edge-capacity constraint vector can be obtained as $\boldsymbol{e}=\left(e_{s_0}\mathbf{1}^{T}_{n_0},\,e_{s_1}\mathbf{1}^{T}_{n_1},\,\ldots,\,e_{s_{k-1}}\mathbf{1}^{T}_{n_{k-1}}\right)^{T}$, where $\mathbf{1}_{n_i}\in\mathbb{R}^{n_i}$ denotes the all-ones vector, and each 
block $e_{s_i}\mathbf{1}^{T}_{n_i}$ corresponds to the edge-capacity limits of all 
ports at layer $i$.

The incidence matrix $M$ is constructed by stacking the port-level 
incidence matrices of each switch layer:
\begin{equation}
\label{tab:heter-inter-M}
M
=
\begin{bmatrix}
    M_{s_0} \\
    M_{s_1} \\
    \vdots \\
    M_{s_{k-1}}
\end{bmatrix},
\end{equation}
where $M_{s_i}\in\{0,1\}^{n_i\times |\mathcal{E}|}$ collects the binary incidence vectors of all 
ports at layer $i$:
\begin{equation}
\label{tab:heter-inter-M_s}
M_{s_i}
=
\begin{bmatrix}
\boldsymbol{m}_{s_ip_1}\\
\vdots\\
\boldsymbol{m}_{s_ip_{n_i}}
\end{bmatrix},
\quad
\boldsymbol{m}_{s_ip_j}\in\{0,1\}^{|\mathcal{E}|},
\end{equation}
where the row vector $\boldsymbol{m}_{s_ip_j}$ indicates which logical edges traverse 
port $j$ at layer $i$: its $l$-th component is equal to one if logical edge $l$ uses 
that port, and zero otherwise. In this way, each row of $M$ encodes a single physical 
switch-port constraint, and the equality $M\mathbf{z}=\boldsymbol{e}$ in 
Eq.~\eqref{tab:heterogeneous-problem-2} enforces the edge-capacity limits of all 
ports across all layers.

\section{Methodology} \label{section5}

% \hy{What is the key contribution of this work? What is the main result? Maybe we should highlight the key points in a remark in this section.} \cz{The solving process is the same as ~\cite{sun2018weighted}, the key contribution is the design of the optimization problem}

To solve the formulated network topology optimization problems, we propose a computationally efficient ADMM-based method as shown in Algorithm~\ref{Algorithm-BA-Topo}. Specifically, we first transform the original optimization problems in \eqref{tab:homogeneous-problem-3} and \eqref{tab:heterogeneous-problem-2} in order to decouple the cardinality constraints from the LMI constraints, and then solve the problems based on the ADMM method with conjugate gradient. 
In this paper, we refer to the topology obtained by solving the optimization problems as BA-Topo (Bandwidth-Aware Topology). The proposed optimization algorithms for homogeneous and heterogeneous bandwidth scenarios are then presented.

\subsection{Homogeneous Bandwidth}

Let $\mathbf{x}=\left[ \mathbf{g}^T,\tilde{\lambda} \right] ^T$, $\mathcal{B} ^+\left( \mathbf{x} \right) =L+\tilde{\lambda} I$, $\mathcal{B} ^- \left( \mathbf{x} \right) =L-\tilde{\lambda} I$, $B_0=\frac{\alpha 11^T}{n}$, and $D\mathbf{x}=\mathbf{diag}\left(L \right) $, where $ D=\left[ \mathrm{abs} \left( A \right) ,\mathbf{0} \right] ,\mathbf{0}\in \mathbb{R} ^n$. Meanwhile, we introduce auxiliary variables $\mathbf{x}_1,S,S_1,\mathbf{y},\mathbf{y}_1,T,T_1$ to decouple the inequality constraints from the equality constraints. To improve readability, we denote $\mathbf{X} =\left[ \mathbf{x}^T,\mathrm{vec}\left( S \right) ^T,\mathbf{y}^T,\mathrm{vec}\left( T \right) ^T \right] ^T $ and $\mathbf{Y} = \left[\mathbf{x}_1^T,\mathrm{vec}\left( S_1 \right) ^T,\mathbf{y}_1^T,\mathrm{vec}\left( T_1 \right) ^T \right] ^T$. Therefore, Eq.~\eqref{tab:homogeneous-problem-3} can be transformed to the following form:
% \begin{equation}
%     \label{tab:homogeneous-problem-4}
%     \begin{matrix}
%         \underset{\mathbf{x}}{\min}&		\boldsymbol{c}^T\mathbf{x}\\
%         s.t.&		\begin{array}{c}
%         \mathbf{x}\ge 0,|\mathbf{x}_{:e}|_0\le r,\\
%         \mathcal{B} ^-\left( \mathbf{x} \right) +B_0\succcurlyeq  \mathbf{0}_{n\times n},\\
%         \mathcal{B} ^+\left( \mathbf{x} \right) \preceq 2 I,\\
%         D\mathbf{x}\leqslant \mathbf{1},\\
%     \end{array}\\
%     \end{matrix}
% \end{equation}
% \hy{(Can we summarize the algorithm into several main steps with subtitles? This can be helpful to show the novelty of the proposed framework.)}
\begin{equation}
    \label{tab:homogeneous-problem-5}
    \begin{matrix}
        &	\underset{\mathbf{X}, \mathbf{Y}}{\min} \,\,	\boldsymbol{c}^T\mathbf{x}\\
        &\\
        s.t.&		\begin{array}{c}
        \mathcal{B} ^-\left( \mathbf{x} \right) +B_0+S=\mathbf{0}_{n\times n},\\
        \mathcal{B} ^+\left( \mathbf{x} \right) +T=2I,\\
        D\mathbf{x}+\mathbf{y}=\mathbf{1},\\
        \mathbf{x}=\mathbf{x}_1,\mathbf{x}_1\geqslant 0,|\left( \mathbf{x}_1 \right) _{:{|\mathcal{E}|}}|_0\leqslant r,\\
        S=S_1,S_1\curlyeqprec \mathbf{0}_{n\times n},\\
        \mathbf{y}=\mathbf{y}_1,\mathbf{y}_1\geqslant 0,T=T_1,T_1\succcurlyeq \mathbf{0}_{n\times n},\\
    \end{array}\\
    \end{matrix}
\end{equation}
where $\boldsymbol{c}\in \mathbb{R} ^{e+1}$ and $\mathbf{x}_{:e}$ represents the first $e$ elements of $\mathbf{x}$. 
The corresponding augmented Lagrangian function is:
\begin{equation}
    \begin{aligned}
    \label{tab:Lagrangian-function}
    \mathcal{L} &=\boldsymbol{c}^T\mathbf{x}+\left< \boldsymbol{\mu },\mathbf{x}-\mathbf{x}_1 \right> +\frac{\rho}{2}\left\| \mathbf{x}-\mathbf{x}_1 \right\| _{F}^{2}
    \\
    &+\left< \mathrm{vec}\left(\varLambda\right) ,\mathrm{vec}\left(S-S_1\right) \right> +\frac{\rho}{2}\left\| S-S_1 \right\| _{F}^{2}
    \\
    &+\left< \boldsymbol{\sigma },\mathbf{y}-\mathbf{y}_1 \right> +\frac{\rho}{2}\left\| \mathbf{y}-\mathbf{y}_{\mathbf{1}} \right\| _{F}^{2}
    \\
    &+\left< \mathrm{vec}\left(\varGamma\right) ,\mathrm{vec}\left(T-T_1\right) \right> +\frac{\rho}{2}\left\| T-T_1 \right\| _{F}^{2},
    \end{aligned}
\end{equation}
where $\boldsymbol{\mu },\varLambda ,\boldsymbol{\sigma },\varGamma$ are dual variables. Using ADMM to solve Eq.~\eqref{tab:homogeneous-problem-5} and setting $\mathbf{D} =\left[ \boldsymbol{\mu }^T,\mathrm{vec}\left( \varLambda \right) ^T,\boldsymbol{\sigma }^T,\mathrm{vec}\left( \varGamma \right) ^T \right] ^T$, we get the following variable update process:

% \yp{(
% Simplify expression?
\begin{equation}
    \begin{aligned}
        \label{tab:ADMM-variable-update}
        \mathbf{Y} ^{k+1}&\coloneqq \mathrm{arg}\min_{\mathbf{Y} \in \mathcal{C} _{\mathbf{Y}}} \mathcal{L} \left( \mathbf{X} ^k, \mathbf{Y} ;\mathbf{D} ^k \right),\\
        \mathbf{X} ^{k+1}&\coloneqq \mathrm{arg}\min_{\mathbf{X} \in \mathcal{C} _{\mathbf{X}}} \mathcal{L} \left( \mathbf{X}, \mathbf{Y} ^{k+1} ;\mathbf{D} ^k \right),\\
        \mathbf{D} ^{k+1}&\coloneqq \mathbf{D} ^k+\rho \left( \mathbf{X}^{k+1}-\mathbf{Y}^{k+1} \right),\\
    \end{aligned}
\end{equation}
% )}
% \begin{equation}
%     \begin{aligned}
%         \label{tab:ADMM-variable-update}
%         \mathbf{X} ^{k+1}&\coloneqq \mathrm{arg}\min_{\mathbf{X} \in \mathcal{C} _{\mathbf{X}}} \mathcal{L} \left( \mathbf{X} ,\mathbf{Y} ^k;\mathbf{D} ^k \right)\\
%         &=\mathrm{arg}\min_{\mathbf{x}_1\in \mathcal{C} _{\mathbf{x}_1}} \mathcal{L} \left( \mathbf{X} ,\mathbf{Y} ^k;\mathbf{D} ^k \right) 
%         \\
%         &+\mathrm{arg}\min_{S_1\in \mathcal{C} _{S_1}} \mathcal{L} _1\left( \mathbf{X} ,\mathbf{Y} ^k;\mathbf{D} ^k \right)\\
%         &+\mathrm{arg}\min_{\mathbf{y}_1\in \mathcal{C} _{\mathbf{y}_1}} \mathcal{L} \left( \mathbf{X} ,\mathbf{Y} ^k;\mathbf{D} ^k \right) 
%         \\
%         &+\mathrm{arg}\min_{T_1\in \mathcal{C} _{T_1}} \mathcal{L} _1\left( \mathbf{X} ,\mathbf{Y} ^k;\mathbf{D} ^k \right)\\
%         \mathbf{Y} ^{k+1}&\coloneqq \mathrm{arg}\min_{\mathbf{Y} \in \mathcal{C} _{\mathbf{Y}}} \mathcal{L} \left( \mathbf{X} ^{k+1},\mathbf{Y} ;\mathbf{D} ^k \right)\\
%         \boldsymbol{\mu }^{k+1}&\coloneqq \boldsymbol{\mu }^k+\rho \left( \mathbf{x}^{k+1}-\mathbf{x}_{1}^{k+1} \right)\\
%         \varLambda ^{k+1}&\coloneqq \varLambda ^k+\rho \left( S^{k+1}-S_{1}^{k+1} \right)\\
%         \boldsymbol{\sigma }^{k+1}&\coloneqq \boldsymbol{\sigma }^k+\rho \left( \mathbf{y}^{k+1}-\mathbf{y}_{1}^{k+1} \right)\\
%         \varGamma ^{k+1}&\coloneqq \varGamma ^k+\rho \left( T^{k+1}-T_{1}^{k+1} \right)\\
%     \end{aligned},
% \end{equation}
where $\mathcal{C} _{\mathbf{x}_1},\mathcal{C} _{S_1},\mathcal{C} _{\mathbf{y}_1},\mathcal{C} _{T_1},\mathcal{C} _{\mathbf{X}},\mathcal{C} _{\mathbf{Y}}$ are feasible domains of variables, defined as follows:
\begin{equation}
    \begin{aligned}
        \label{tab:feasible_region}
\mathcal{C} _{\mathbf{x}_1}&\coloneqq \left\{ \mathbf{x}_1|\mathbf{x}_1\geqslant 0,|\left( \mathbf{x}_1 \right) _{:e}|_0\leqslant r \right\}, 
\\
\mathcal{C} _{S_1}&\coloneqq \left\{ S_1|S_1\curlyeqprec \mathbf{0}_{n\times n} \right\}, 
\\
\mathcal{C} _{\mathbf{y}_1}&\coloneqq \left\{ \mathbf{y}_1|\mathbf{y}_1\geqslant 0 \right\}, 
\\
\mathcal{C} _{T_1}&\coloneqq \left\{ T_1|T_1\succcurlyeq \mathbf{0}_{n\times n} \right\}, 
\\
\mathcal{C} _{\mathbf{Y}}&\coloneqq \mathcal{C} _{\mathbf{x}_1} \otimes  \mathcal{C} _{S_1} \otimes \mathcal{C} _{\mathbf{y}_1} \otimes \mathcal{C} _{T_1},
\\
\mathcal{C} _{\mathbf{X}}&\coloneqq \left\{ \mathbf{x},S,\mathbf{y},T\left| \begin{array}{c}
	\mathcal{B} ^-\left( \mathbf{x} \right) +B_0+S=\mathbf{0}_{n\times n},\\
	\mathcal{B} ^+\left( \mathbf{x} \right) +T=2I,\\
	D\mathbf{x}+\mathbf{y}=\mathbf{1}\\
\end{array} \right. \right\}.       
    \end{aligned}
\end{equation}

The updating of $\mathbf{x}_1,S_1,\mathbf{y}_1,T_1$ can be done separately, so the following update process can be obtained:
\begin{equation}
    \label{tab:update-of-auxiliary-variables}
    \begin{aligned}
        % \mathbf{x}_{1}^{k+1}&=\mathrm{Proj}_{\left\{ \mathbf{x}_1||\left( \mathbf{x}_1 \right) _{:e}|_0\le r \right\}}\left( \mathrm{Proj}_{\left\{ \mathbf{x}_1|\mathbf{x}_1\geqslant 0 \right\}}\left( \mathbf{x}^k+\frac{\boldsymbol{\mu }^k}{\rho} \right) \right) 
        % \\
        % S_{1}^{k+1}&=\mathrm{Proj}_{\left\{ S_1|S_1\preceq \mathbf{0}_{n\times n} \right\}}\left( S^k+\frac{\varLambda ^k}{\rho} \right) 
        % \\
        % \mathbf{y}_{1}^{k+1}&=\mathrm{Proj}_{\left\{ \mathbf{y}_1|\mathbf{y}_1\geqslant 0 \right\}}\left( \mathbf{y}^k+\frac{\boldsymbol{\mu }^k}{\rho} \right) 
        % \\
        % T_{1}^{k+1}&=\mathrm{Proj}_{\left\{ T_1|T_1\succcurlyeq \mathbf{0}_{n\times n} \right\}}\left( T^k+\frac{\varGamma ^k}{\rho} \right) 
        % \\
        \mathbf{x}_{1}^{k+1}&=\mathrm{Proj}_{\mathcal{C}_{\mathbf{x}_1}}\left( \mathbf{x}^k+\frac{\boldsymbol{\mu }^k}{\rho} \right),  
        S_{1}^{k+1}=\mathrm{Proj}_{\mathcal{C} _{S_1}}\left( S^k+\frac{\varLambda ^k}{\rho} \right), 
        \\
        \mathbf{y}_{1}^{k+1}&=\mathrm{Proj}_{\mathcal{C} _{\mathbf{y}_1}}\left( \mathbf{y}^k+\frac{\boldsymbol{\sigma }^k}{\rho} \right),  
        T_{1}^{k+1}=\mathrm{Proj}_{\mathcal{C} _{T_1}}\left( T^k+\frac{\varGamma ^k}{\rho} \right).
    \end{aligned}
\end{equation}
For $\mathbf{x}_1\geqslant 0$ and $\mathbf{y}_1\geqslant 0$, we project by keeping the non-negative elements and setting the others to zero. For $|\left( \mathbf{x}_1 \right) _{:{|\mathcal{E}|}}|_0\leqslant r$, we project by keeping the largest $r$ elements of the first ${|\mathcal{E}|}$ elements of $\mathbf{x}_1$ and setting the others to zero. For $S_1\preceq \mathbf{0}_{n\times n}$, we project by the following equation:
\begin{equation}
    \label{tab:projection-S1}
    S_1=U\mathbf{diag}\left( \min \left( \mathbf{diag}\left( \varSigma \right) ,0 \right) \right) U^T,
\end{equation}
where $U$ and $\varSigma$ are the matrices obtained by eigenvalue decomposition of $S_1$. For $T_1\succcurlyeq \mathbf{0}_{n\times n}$, the projection method is the same as that of Eq.~\eqref{tab:projection-S1}.

For $\mathcal{C} _{\mathbf{Y}}$ in Eq.~\eqref{tab:feasible_region}, it can be viewed as the linear equation constraints $A\mathbf{X}=\boldsymbol{b}$,
% \begin{equation}
%     % A\left( \mathbf{x}^T,\mathrm{vec}\left( S \right) ^T,\mathbf{y}^T,\mathrm{vec}\left( T \right) ^T \right) ^T=\boldsymbol{b},
%     A\mathbf{X}=\boldsymbol{b},
% \end{equation}
where A and b are defined as follows:
\begin{equation}
    \begin{aligned}
        A&=\left[ \begin{matrix}
	\begin{array}{c}
	\tilde{B}^-\\
	\tilde{B}^+\\
	D\\
\end{array}&		\begin{array}{c}
	I_{n^2}\\
	\mathbf{0}_{n^2\times n^2}\\
	\mathbf{0}_{n\times n^2}\\
\end{array}&		\begin{array}{c}
	\mathbf{0}_{n^2\times n}\\
	\mathbf{0}_{n^2\times n}\\
	I_n\\
\end{array}&		\begin{array}{c}
	\mathbf{0}_{n^2\times n^2}\\
	I_{n^2}\\
	\mathbf{0}_{n\times n^2}\\
\end{array}\\
\end{matrix} \right] ,
\\
\boldsymbol{b}&=\left[ \mathrm{vec}\left( -B_0 \right) ^T,\mathrm{vec}\left( 2I \right) ^T,\mathbf{1}^T \right] ^T,
    \end{aligned}
\end{equation}
where $\tilde{B}^-\mathbf{x}=\mathrm{vec}\left( \mathcal{B} ^-\left( \mathbf{x} \right) \right),\tilde{B}^+\mathbf{x}=\mathrm{vec}\left( \mathcal{B} ^+\left( \mathbf{x} \right) \right)$. Therefore, we can update $\mathbf{x},S,\mathbf{y},T$ by solving the following linear equations:
\begin{equation}
    \label{tab:update-primal-variables-homogenous}
    \begin{aligned}
\underset{\tilde{A}}{\underbrace{\left[ \begin{matrix}
	I&		A^T\\
	A&		0\\
\end{matrix} \right] }}\left[ \begin{array}{c}
	\mathbf{X}^{k+1}\\
	\boldsymbol{\lambda }\\
\end{array} \right] =\underset{\tilde{b}}{\underbrace{\left[ \begin{array}{c}
	\mathbf{Y}^{k+1}-\frac{\mathbf{D}^k+C}{\rho}\\
	\boldsymbol{b}\\
\end{array} \right] }},
    \end{aligned}
\end{equation}
where $\boldsymbol{\lambda }$ is the dual variable associated with the linear equality constraint, and $C=\left[ \boldsymbol{c}^T, \mathbf{0}^T \right] ^T$ with proper dimension.

% $I$ and $\mathbf{0}$ are square matrices with dimensions corresponding to the number of columns and rows of matrix $A$, respectively,

\subsection{Heterogeneous Bandwidth}
% The above is the solving process of the topology optimization problem in homogeneous bandwidth. 
Similar with the case of homogeneous bandwidth, by introducing auxiliary variables $\mathbf{x}_1,S,S_1,\mathbf{y},\mathbf{y}_1,T,T_1,\mathbf{z},\mathbf{z_1},\mathbf{\nu },\mathbf{\nu }_1$ and letting
% $\mathbf{X}^{\prime} =\left[ \mathbf{x}^T, \mathrm{vec}\left( S \right) ^T, \mathbf{y}^T, \mathrm{vec}\left( T \right) ^T, \mathbf{z}^T, \mathbf{\nu }^T \right] ^T$, 
$\mathbf{X}^{\prime} =\left[ \mathbf{X}^T, \mathbf{z}^T, \mathbf{\nu }^T \right] ^T$, 
% $\mathbf{Y}^{\prime} =\left[ \mathbf{x}_1^T, \mathrm{vec}\left(S_1\right)^T, \mathbf{y}_1^T, \mathrm{vec}\left(T_1\right)^T, \mathbf{z}_{1}^T, \mathbf{\nu }_{1}^T \right] $
$\mathbf{Y}^{\prime} =\left[ \mathbf{Y}^T, \mathbf{z}_{1}^T, \mathbf{\nu }_{1}^T \right] $. Then, the optimization problem in \eqref{tab:heterogeneous-problem-2} becomes
\begin{equation}
    \label{tab:heterogeneous-problem-3}
    % \begin{matrix}
    %     &	\underset{\mathbf{X},\mathbf{Y}}{\min} \,\,	\boldsymbol{c}^T\mathbf{x}\\
    %     s.t.&		\begin{array}{c}
    %     \mathcal{B} ^-\left( \mathbf{x} \right) +B_0+S=\mathbf{0}_{n\times n},\\
    %     \mathcal{B} ^+\left( \mathbf{x} \right) +T=2I,\\
    %     D\mathbf{x}+\mathbf{y}=\mathbf{1},\\
    %     M\mathbf{z}=\boldsymbol{e},\,\,
    %     \mathbf{g}-\mathbf{z}+\mathbf{\nu }=\mathbf{0},\\
    %     \mathbf{x}=\mathbf{x}_1,\mathbf{x}_1\ge 0,\\
    %     S=S_1,S_1\preceq \mathbf{0}_{n\times n},\\
    %     \mathbf{y}=\mathbf{y}_1,\mathbf{y}_1\ge 0,T=T_1,T_1\succcurlyeq \mathbf{0}_{n\times n},\\
    %     \mathbf{z}=\mathbf{z}_1,\mathbf{z}_1\in \left\{ 0,1 \right\} ^m,\mathbf{\nu }=\mathbf{\nu }_1,\mathbf{\nu }_1\ge 0.\\
    % \end{array}\\
    % \end{matrix}
\begin{matrix}
	&		\underset{\mathbf{X}^{\prime},\mathbf{Y}^{\prime}}{\min}\,\,\boldsymbol{c}^T\mathbf{x}\\
 &\\
	s.t.&		\begin{array}{c}
	\left\{ \mathbf{x},S,\mathbf{y},T \right\} \in \mathcal{C} _{\mathbf{X}}, \left\{ \mathbf{x}_1,S_1,\mathbf{y}_1,T_1 \right\} \in \mathcal{C} _{\mathbf{Y}}\\
	\mathbf{x}=\mathbf{x}_1, S=S_1, \mathbf{y}=\mathbf{y}_1, T=T_1,
	\\
	M\mathbf{z}=\boldsymbol{e},\,\,\mathbf{g}-\mathbf{z}+\mathbf{\nu }=0,\\
	\mathbf{z}=\mathbf{z}_1,\mathbf{z}_1\in \left\{ 0,1 \right\} ^{|\mathcal{E}|},\mathbf{\nu }=\mathbf{\nu }_1,\mathbf{\nu }_1\geqslant 0.\\
\end{array}\\
\end{matrix}
\end{equation}
The corresponding augmented Lagrangian function is:
\begin{equation}
    \begin{aligned}
    \label{tab:augmented-Lagrangian-heterogeneous}
    \mathcal{L}^{\prime} &= \mathcal{L}
    % \boldsymbol{c}^T\mathbf{x}+\left< \boldsymbol{\mu },\mathbf{x}-\mathbf{x}_1 \right> +\frac{\rho}{2}\left\| \mathbf{x}-\mathbf{x}_1 \right\| _{F}^{2}
    % \\
    % &+\left< \mathrm{vec}\left(\varLambda\right) ,\mathrm{vec}\left(S-S_1\right) \right> 
    % +\frac{\rho}{2}\left\| S-S_1 \right\| _{F}^{2}
    % \\
    % &+\left< \boldsymbol{\sigma },\mathbf{y}-\mathbf{y}_1 \right> +\frac{\rho}{2}\left\| \mathbf{y}-\mathbf{y}_1 \right\| _{F}^{2}
    % \\
    % &+\left< \mathrm{vec}\left(\varGamma\right) ,\mathrm{vec}\left(T-T_1\right) \right> +\frac{\rho}{2}\left\| T-T_1 \right\| _{F}^{2}
    +\left< \boldsymbol{\iota },\mathbf{z}-\mathbf{z}_1 \right> + \frac{\rho}{2}\left\| \mathbf{z}-\mathbf{z}_1 \right\| _{F}^{2}
    \\
    &~~~+\left< \boldsymbol{\kappa },\mathbf{\nu }-\mathbf{\nu }_1 \right> + \frac{\rho}{2}\left\| \mathbf{\nu }-\mathbf{\nu }_1 \right\| _{F}^{2}.
    \end{aligned}
\end{equation}
Let $\mathbf{D}^{\prime} =\left[ \mathbf{D}^T, \boldsymbol{\iota}^T, \boldsymbol{\kappa }^T \right] ^T $. Then, the update of the variables $\mathbf{Y}^{\prime}$ becomes:
\begin{equation}
    \label{tab:update-process-auxiliary-variables-heter}
    \begin{aligned}
        \mathbf{Y}^{\prime k+1}&=\mathrm{Proj}\left( \mathbf{X}^{\prime k}+\frac{\mathbf{D}^{\prime k}}{\rho} \right),
    \end{aligned}
\end{equation}
where the projection of each variable corresponds to its own constraint space which is omit for brevity.

For $\mathbf{z}_1\in \left\{ 0,1 \right\} ^{|\mathcal{E}|}$, considering that the result after projection needs to satisfy the constraint on the number of edges, we project by setting the largest $r$ element in $\mathbf{z}_1$ to one, and setting the other elements to zero.

Similar to the homogeneous case, 
% the update of the variables set $\mathbf{X}$ is obtained by solving Eq. \ref{tab:update-primal-variables-homogenous} with $A$, $\mathbf{b}$ and $C$ redefined as follows:
the update of the variables $\mathbf{X}^{\prime}$ is obtained by solving the following equation:
\begin{equation}
    \label{tab:update-process-primal-variables-heter}
\underset{\tilde{A}^{\prime}}{\underbrace{\left[ \begin{matrix}
	I&		A^{\prime T}\\
	A^{\prime}&		0\\
\end{matrix} \right] }}\left[ \begin{array}{c}
	\mathbf{X}^{\prime k+1}\\
	\boldsymbol{\lambda }^{\prime}\\
\end{array} \right] =\underset{\tilde{b}^{\prime}}{\underbrace{\left[ \begin{array}{c}
	\mathbf{Y}^{\prime k+1}-\frac{\mathbf{D}^{\prime k}+C^{\prime}}{\rho}\\
	\boldsymbol{b}^{\prime}\\
\end{array} \right] }},
\end{equation}
where $A^{\prime}$, $\boldsymbol{b}^{\prime}$ and $C^{\prime}$ are defined as follows:
\begin{equation}
\small
    \begin{aligned}
%         A&=\left[ \begin{matrix}
% 	\begin{array}{l}
% 	\tilde{B}^-\\
% 	\tilde{B}^+\\
% 	D\\
% 	\mathbf{0}_{m\times \left( e+1 \right)}\\
% 	I\_\\
% \end{array}&		\begin{array}{c}
% 	I_{n^2}\\
% 	\mathbf{0}_{n^2\times n^2}\\
% 	\mathbf{0}_{n\times n^2}\\
% 	\mathbf{0}_{m\times n^2}\\
% 	\mathbf{0}_{e\times n^2}\\
% \end{array}&		\begin{array}{c}
% 	\mathbf{0}_{n^2\times n}\\
% 	\mathbf{0}_{n^2\times n}\\
% 	I_n\\
% 	\mathbf{0}_{m\times n}\\
% 	\mathbf{0}_{e\times n}\\
% \end{array}&		\begin{array}{c}
% 	\mathbf{0}_{n^2\times n^2}\\
% 	I_{n^2}\\
% 	\mathbf{0}_{n\times n^2}\\
% 	\mathbf{0}_{m\times n^2}\\
% 	\mathbf{0}_{e\times n^2}\\
% \end{array}&		\begin{array}{c}
% 	\mathbf{0}_{n^2\times e}\\
% 	\mathbf{0}_{n^2\times e}\\
% 	\mathbf{0}_{n\times e}\\
% 	M\\
% 	-I_e\\
% \end{array}&		\begin{array}{c}
% 	\mathbf{0}_{n^2\times e}\\
% 	\mathbf{0}_{n^2\times e}\\
% 	\mathbf{0}_{n\times e}\\
% 	\mathbf{0}_{m\times e}\\
% 	I_e\\
% \end{array}\\
% \end{matrix} \right] ,
% \\
% \boldsymbol{b}&=\left( \mathrm{vec}\left( -B_0 \right) ^T,\mathrm{vec}\left( 2I \right) ^T,\mathbf{1}^T,\boldsymbol{e}^T,\mathbf{0}^T \right) ^T,
A^{\prime}&=\left[ \begin{matrix}
	\tilde{B}^-&		I_{n^2}&		\mathbf{0}_{n^2\times n}&		\mathbf{0}_{n^2\times n^2}&		\mathbf{0}_{n^2\times {|\mathcal{E}|}}&		\mathbf{0}_{n^2\times {|\mathcal{E}|}}\\
	\tilde{B}^+&\mathbf{0}_{n^2\times n^2}&\mathbf{0}_{n^2\times n}&I_{n^2}&\mathbf{0}_{n^2\times {|\mathcal{E}|}}&\mathbf{0}_{n^2\times {|\mathcal{E}|}}\\
	D&\mathbf{0}_{n\times n^2}&I_n&\mathbf{0}_{n\times n^2}&\mathbf{0}_{n\times {|\mathcal{E}|}}&\mathbf{0}_{n\times {|\mathcal{E}|}}\\
	\mathbf{0}_{q\times \left( {|\mathcal{E}|}+1 \right)}&\mathbf{0}_{q\times n^2}&\mathbf{0}_{q\times n}&\mathbf{0}_{q\times n^2}&M&\mathbf{0}_{q\times {|\mathcal{E}|}}\\
	I\_&\mathbf{0}_{{|\mathcal{E}|}\times n^2}&\mathbf{0}_{{|\mathcal{E}|}\times n}&\mathbf{0}_{{|\mathcal{E}|}\times n^2}&-I_{|\mathcal{E}|}&I_{|\mathcal{E}|}\\
\end{matrix} \right],
\\
\boldsymbol{b}^{\prime}&=\left[ \mathrm{vec}\left( -B_0 \right) ^T,\mathrm{vec}\left( 2I \right) ^T,\mathbf{1}^T,\boldsymbol{e}^T,\mathbf{0}^T \right] ^T,
\\
C^{\prime}&=\left[ \boldsymbol{c}^T, \mathbf{0}^T \right] ^T,~\text{with}~ I\_=\left[ I_{|\mathcal{E}|},\mathbf{0}_{|\mathcal{E}|} \right].
    \end{aligned}
\end{equation}

The update of the dual variables $\mathbf{D}^{\prime}$ is as follows:
\begin{equation}
    \label{tab:dual-variables-update-heter}
    \begin{aligned}
 %        \boldsymbol{\mu }^{k+1}&\coloneqq \boldsymbol{\mu }^k+\rho \left( \mathbf{x}^{k+1}-\mathbf{x}_{1}^{k+1} \right)\\
	% \varLambda ^{k+1}&\coloneqq \varLambda ^k+\rho \left( S^{k+1}-S_{1}^{k+1} \right)\\
	% \boldsymbol{\sigma }^{k+1}&\coloneqq \boldsymbol{\sigma }^k+\rho \left( \mathbf{y}^{k+1}-\mathbf{y}_{1}^{k+1} \right)\\
	% \varGamma ^{k+1}&\coloneqq \varSigma ^k+\rho \left( T^{k+1}-T_{1}^{k+1} \right)\\
	% \boldsymbol{\iota }^{k+1}&\coloneqq \boldsymbol{\iota }^k+\rho \left( \mathbf{z}^{k+1}-\mathbf{z}_{1}^{k+1} \right)\\
	% \boldsymbol{\kappa }^{k+1}&\coloneqq \boldsymbol{\kappa }^k+\rho \left( \mathbf{\nu }^{k+1}-\mathbf{\nu }_{1}^{k+1} \right)\\
    \mathbf{D}^{\prime k+1}&\coloneqq \mathbf{D}^{\prime k}+\rho \left( \mathbf{X}^{\prime k+1}-\mathbf{Y}^{\prime k+1} \right).
    \end{aligned}
\end{equation}

\subsection{Computation Acceleration}
Algorithm 2 consists of several sub-steps that are straightforward to solve, except for the step involving the update of variables in $Y$, which requires solving a large linear system of equations (c.f., Eq.~\eqref{tab:update-primal-variables-homogenous} and Eq.~\eqref{tab:update-process-primal-variables-heter}). Noticing that direct inversion of the coefficient matrix is not feasible for such large-scale systems, we employ the Bi-Conjugate Gradient Stabilized Method (Bi-CGSTAB)~\cite{van1992bi} to solve it iteratively (c.f., Line 6 and 15 in Algorithm \ref{Algorithm-BA-Topo}). This algorithm is based on the biconjugate gradient method (BiCG), which uses two vector sequences to ensure that the residual vector and direction vector generated in each iteration are biconjugate with each other. Also, Bi-CGSTAB introduces stabilization steps to enhance numerical stability. Therefore, this method can  efficiently and stably solve the linear system with the two indefinite coefficient matrices in Eq.~\eqref{tab:update-primal-variables-homogenous} and Eq.~\eqref{tab:update-process-primal-variables-heter}.

Moreover, since the coefficient matrix remains constant throughout the iterations of the ADMM algorithm, we precompute its incomplete LU decomposition (ILU)~\cite{meijerink1977iterative} during the initialization phase (c.f., Line 3 and 12 in Algorithm \ref{Algorithm-BA-Topo}). This precomputed ILU is used as a preconditioner for the Bi-CGSTAB method to accelerate convergence. Furthermore, given that the coefficient matrix is sparse, we employ Compressed Sparse Column (CSC) storage~\cite{davis2006direct} to optimize computation and memory usage. As a result, our algorithm successfully scales to hundreds of nodes, making it suitable for most decentralized learning scenarios.

\begin{center}
\begin{algorithm}[!htbp]
        \caption{ADMM-Based Framework for Solving the Network Topology Optimization Problems} 
    %   \small
        \label{Algorithm-BA-Topo} 
        \begin{algorithmic}[1]
        \Require{$n,r,\rho,\alpha$, initial point $\mathbf{x}^0,\mathbf{x}_{1}^{0},S^0,S_{1}^{0},\mathbf{y}^0, \mathbf{y}_{1}^{0}, T^0,T_{1}^{0},$ $ \boldsymbol{\mu }^0, \varLambda ^0, \boldsymbol{\sigma }^0, \varGamma ^0$,  algorithm convergence error $\epsilon$, initial point for heterogeneous bandwidth scenario $\mathbf{z}^0,\mathbf{z}_{1}^{0},\mathbf{\nu }^0,\mathbf{\nu }_{1}^{0},\boldsymbol{\iota }^0,\boldsymbol{\kappa }^0$}.
        \Ensure{Homogeneous bandwidth: locally optimal solution of Eq.~\eqref{tab:homogeneous-problem-5}, $\mathbf{x},\mathbf{x}_1,S,S_1,\mathbf{y},\mathbf{y}_1,T,T_1$; heterogeneous bandwidth: locally optimal solution of Eq.~\eqref{tab:heterogeneous-problem-3}, $\mathbf{x},\mathbf{x}_1,S,S_1,\mathbf{y},\mathbf{y}_1,T,T_1,\mathbf{z},\mathbf{z}_1,\mathbf{\nu },\mathbf{\nu }_1$}.
        \State Set $k=0$.
        \If{the optimization problem belongs to the homogeneous bandwidth scenario:}
        \State compute $M$ = ILU($\tilde{A}$).
        \While{$\left\| \mathbf{x}^k-\mathbf{x}_{1}^{k} \right\| _{F}^{2}+\left\| S^k-S_{1}^{k} \right\| _{F}^{2}+\left\| \mathbf{y}^k-\mathbf{y}_{1}^{k} \right\| _{F}^{2}+\left\| T^k-T_{1}^{k} \right\| _{F}^{2}>\epsilon $}
            \State Update $\mathbf{x}_1,S_1,\mathbf{y}_1,T_1$ based on Eq.~\eqref{tab:update-of-auxiliary-variables}.
            \State Update $\mathbf{x},S,\mathbf{y},T$ by BI-CGSTAB($\tilde{A}$, $\tilde{b}$, $M$).
            \State Update $\boldsymbol{\mu },\varLambda ,\boldsymbol{\sigma },\varGamma$ based on Eq.~\eqref{tab:ADMM-variable-update}.
            \State Set $k=k+1$.
        \EndWhile
        \State Determine $\mathbf{x},\mathbf{x}_1,S,S_1,\mathbf{y},\mathbf{y}_1,T,T_1$.
        \Else
        \State compute $M$ = ILU($\tilde{A}^{\prime}$).
        \While{$\left\| \mathbf{x}^k-\mathbf{x}_{1}^{k} \right\| _{F}^{2}+\left\| S^k-S_{1}^{k} \right\| _{F}^{2}+\left\| \mathbf{y}^k-\mathbf{y}_{1}^{k} \right\| _{F}^{2}+\left\| T^k-T_{1}^{k} \right\| _{F}^{2}+\left\| \mathbf{z}^k-\mathbf{z}_{1}^{k} \right\| _{F}^{2}+\left\| \mathbf{\nu }^k-\mathbf{\nu }_{1}^{k} \right\| _{F}^{2}>\epsilon $}
            \State Update $\mathbf{x}_1,S_1,\mathbf{y}_1,T_1,\mathbf{z}_1,\mathbf{\nu }_1$ based on Eq.~\eqref{tab:update-process-auxiliary-variables-heter}.
            \State  Update $\mathbf{x},S,\mathbf{y},T,\mathbf{z},\mathbf{\nu }$ by BI-CGSTAB($\tilde{A}^{\prime}$, $\tilde{b}^{\prime}$, $M$).
            \State Update $\boldsymbol{\mu },\varLambda ,\boldsymbol{\sigma },\varGamma,\boldsymbol{\iota },\boldsymbol{\kappa }$ based on Eq.~\eqref{tab:dual-variables-update-heter}.
            \State Set $k=k+1$.
        \EndWhile
        \State Determine $\mathbf{x},\mathbf{x}_1,S,S_1,\mathbf{y},\mathbf{y}_1,T,T_1,\mathbf{z},\mathbf{z}_1,\mathbf{\nu },\mathbf{\nu }_1$.
        \EndIf 
        
        \end{algorithmic}
\end{algorithm}
% \end{minipage}
\end{center}
\begin{remark}
    % To deal with the cardinality constraints, we introduce auxiliary variables to decouple them from the LMI constraints, thus making it amenable to be solved under the ADMM-based framework.
    Compared to the formulated optimization problems in \eqref{tab:homogeneous-problem-3} and \eqref{tab:heterogeneous-problem-2}, the equivalent Mixed-Integer SDP problems in \eqref{tab:homogeneous-problem-5} and \eqref{tab:heterogeneous-problem-3} can be efficiently solved iteratively by our proposed ADMM-based Algorithm \ref{Algorithm-BA-Topo}.
    More importantly, compared with the existing methods, the proposed network topology optimization framework can obtain topologies in the full solution space rather than in the subset of the solution space~\cite{xiao2004fast}, leading to faster consensus and better scalability, as will be shown in the experiments.
\end{remark}

% \yp{
% \begin{remark}
% Since the optimization is sensitive to initialization, we generate the initial topology using simulated annealing~\cite{kirkpatrick1983optimization} so that it has a small average shortest path length (ASPL), a metric known to correlate with low communication delay in networked systems~\cite{koibuchi2016optical}. Such initialization helps the solver avoid poor local optima and improves the chance of obtaining topologies with faster consensus.
% \end{remark}
% }

% \yp{
% \begin{remark}
%     For time-invariant static physical topology, it is only necessary to calculate the topology optimization results and distribute the weight matrix to all nodes during initialization. The entire training process is still distributed and decentralized.
% \end{remark}
% }

\section{Experiment Results} \label{section6}
% Experimental results in different scenarios 
In this section, we verify the effectiveness of the proposed network topology optimization method in terms of consensus speed (c.f., Sec. \ref{consensus-experiments}) and efficiency of decentralized learning tasks (c.f., Sec. \ref{decentralized-experiments}). The topologies considered in the experiments include ring, 2D grid~\cite{nedic2018network}, 2D torus ~\cite{nedic2018network}, exponential ~\cite{ying2021exponential}, and Equitopo~\cite{song2022communication}, which covers both classic topologies and SOTA topology. For a fair comparison, we chose the static undirected graph U-EquiStatic of the same type as BA-Topo for the four variants of Equitopo.
We conduct the experiments on the hardware system with 2 Intel Xeon Gold 6226R processors and 8 NVIDIA GeForce RTX 2080 Ti with 11 GB memory per GPU. All the experiments are implemented using pytorch 2.0.0 with gloo as the communication backend.

It should be noted that as the optimization problem is sensitive to initialization, we construct the initial topology using simulated annealing~\cite{kirkpatrick1983optimization} to yield a small average shortest path length (ASPL), a metric known to correlate with low communication delay in networked systems~\cite{koibuchi2016optical}. This warm-start strategy helps avoid poor local optima and increases the likelihood of obtaining topologies with faster consensus.
% First, we compare the consensus speed of BA-Topo with existing communication topologies in Sec.~\ref{consensus-experiments}, and verify that BA-Topo can achieve faster consensus speed. Then, we verify the effectiveness of BA-Topo for decentralized learning tasks in Sec.~\ref{decentralized-experiments}, showing that BA-Topo outperforms existing communication topologies. 

% \subsection{Bandwidth heterogeneous scenarios setting} \label{consensus-experiments}

% \subsubsection{Inter-Server Switch Port Bandwidth Heterogeneity}
\subsection{Consensus Speed} \label{consensus-experiments}
In order to measure the consensus speed of network topologies, we first randomly initialize $\boldsymbol{x}_{i,0}$ for node $i$ with standard Gaussian distribution. Then, we update $\boldsymbol{x}_{i,k}$ by $\boldsymbol{x}_{i,k}=W_{ii}\boldsymbol{x}_{i,k-1}+\sum_{j\in \mathcal{N} _i}{W_{ij}\boldsymbol{x}_{j,k-1}}$ and evaluate the consensus error $\left\| \boldsymbol{x}_k-\bar{\boldsymbol{x}} \right\| _2$ over time.
% Since the bandwidth of the edge is closely related to the underlying physical structure of the hardware system, the available bandwidth of each node cannot be configured as required. Therefore, we need to evaluate the time per iteration by simulation
To evaluate the time required for reaching consensus,
we first obtain the available bandwidth of an edge in the server as 9.76 GB/s (denoted as $b_{avail}$) through actual measurements~\cite{pei2019iteration, barrachina2023using}, and use this as the baseline for the maximum available bandwidth of each node.
Subsequently, we test the communication time for communicating the initial model parameters $\boldsymbol{x}_{i,0}$ on the link with a bandwidth of 9.76GB/s, which is 5.01ms (denoted as $t_{comm}$).
Then, the smallest available bandwidth in all edges is calculated to scale the iteration time:
\begin{equation}
        \label{tab:communication-time-scale}
        % t_{iter}=\frac{9.76}{b_{min}}\times 5.01.
        t_{iter}=\frac{b_{avail}}{b_{min}}\times {t_{comm}},
\end{equation}
where $b_{min}$ represents the minimal bandwidth of all edges.

% \begin{table*}[t]
%     \center
%     \caption{Experimental parameters.}
%     \label{tab:iso-bandwidth-consensus}
%     \setlength{\tabcolsep}{1.6mm} %可随机设置，调整表格整体长度到适合自己的大小为止
%     \scalebox{1}{
%     \begin{tabular}{c|c}
%       \toprule[2pt]
%       % \multicolumn{2}{c|}{Parameter}
%       \multirow[t]{2}{*}{$b_{avail}$} & 9.76\\ \midrule[1pt]
%       \multirow[t]{2}{*}{$t_{comm}$} & 5.01\\ \midrule[1pt]
%       \multirow[t]{2}{*}{$t_{comp}$} & 15.21\\ \bottomrule[2pt]
%     \end{tabular}
%     }
% \end{table*}

\subsubsection{Homogeneous Bandwidth} \label{homo-consensus}

% \arrayrulecolor{blue}
\begin{table*}[t]
    \center
    \caption{Comparison of asymptotic convergence factor and convergence time across different numbers of nodes.}
    \label{tab:iso-bandwidth-consensus}
    \setlength{\tabcolsep}{1.6mm}{ %可随机设置，调整表格整体长度到适合自己的大小为止
    \scalebox{1}{
    \begin{tabular}{c|c|cccccccccccccc}
      \toprule[2pt]
      \multicolumn{2}{c|}{Number of Nodes}            & 4 & 6 & 8 & 12 & 16 & 24 & 32 & 48 & 64 & 96 & 128 \\ \midrule[1pt]
      \multirow{2}{*}{Asymptotic Convergence Factor} & exponential & \textbf{0.33} & 0.5 & 0.5 & 0.6 & 0.6 & 0.67 & 0.67 & 0.71 & 0.71 & 0.75 & 0.75 \\ %\cline{2-16}
                                    & U-EquiStatic (EquiTopo) & 0.5 &  \textbf{0.25} & 0.43 & 0.54 & \textbf{0.47} & 0.54 & 0.62 & 0.68 & 0.74 & 0.67 & 0.72 \\ 
                                   & BA-Topo  & \textbf{0.33} & 0.33 & \textbf{0.41} & \textbf{0.5} & 0.52 & \textbf{0.51} & \textbf{0.54} & \textbf{0.55} & \textbf{0.57} & \textbf{0.61} & \textbf{0.67} \\ \midrule[1pt]
      \multirow{2}{*}{Convergence Time/ms}     & exponential & \textbf{90} & 210 & 210 & 381 & 381 & 576 & 576 & 842 & 842 & 1157 & 1157 \\ %\cline{2-16}
                                   & U-EquiStatic (EquiTopo) & 140 & \textbf{140} & 240 & 341 & 391 & \textbf{481} & 631 & 782 & 992 & 1002 & 1242 \\ 
                                   & BA-Topo & \textbf{90} & 150 & \textbf{180} & \textbf{301} & \textbf{351} & \textbf{481} & \textbf{541} & \textbf{631} & \textbf{762} & \textbf{992} & \textbf{1127} \\ \bottomrule[2pt]
    \end{tabular}
    \vspace{-0.8cm}
    }}
\end{table*}
% \arrayrulecolor{black}

\begin{figure}[!htpb]
  \centering
  \includegraphics[width=0.85\linewidth]{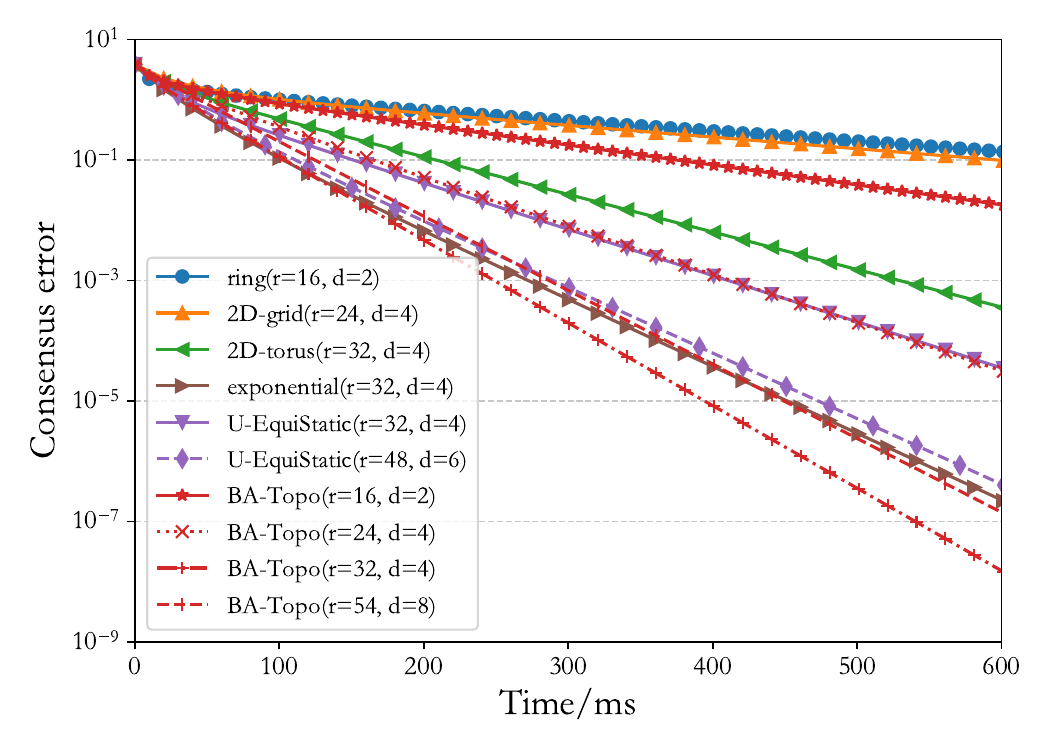}
  \caption{Comparison of consensus speed among various topologies with $n=16$ in homogeneous bandwidth scenario.}
  \label{fig:consensus_error_homo}
\end{figure}

We set the number of nodes to 16 and the available bandwidth per node to 9.76GB/s. The available bandwidth on edge $\left\{ i,j \right\} $ is given by $\min \left( \frac{9.76}{d_i},\frac{9.76}{d_j} \right) $, where $d_i$ and $d_j$ are the degrees of nodes $i$ and $j$ (in exponential topology, this refers to the out-degree/in-degree of nodes). Fig~\ref{fig:consensus_error_homo} shows how the consensus errors on different topologies change over time, where $r$ is the number of edges and $d$ is the maximum degree of the topology. It can be seen that the consensus speed of BA-Topo outperforms the existing network topologies when the number of edges is the same, and BA-Topo has the fastest consensus speed among all topologies when the number of edges is 32. 

\subsubsection{Node-Level Bandwidth Heterogeneity} \label{Node-Level Bandwidth Heterogeneity}
\begin{figure}[!htpb]
  \centering
  \includegraphics[width=0.85\linewidth]{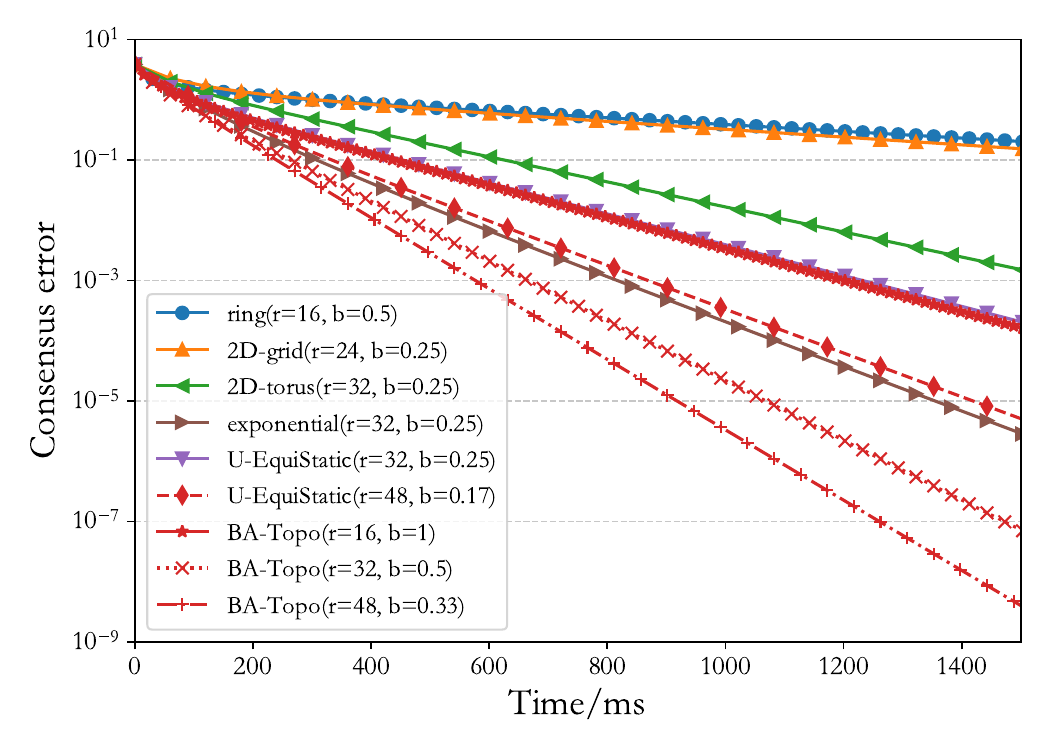}
  \caption{Comparison of consensus speed among various topologies with $n=16$ in node-level bandwidth heterogeneity scenario.}
  \label{fig:consensus_error_heter}
\end{figure}

We set the number of nodes to 16, with the bandwidth ratios for the nodes being 3:3:3:3:3:3:3:3:1:1:1:1:1:1:1:1. The bandwidth of node 1 to node 8 is 9.76GB/s, while the bandwidth of node 9 to node 16 is 3.25GB/s. The available bandwidth on edge $\left\{ i,j \right\} $ is given by $\min \left( \frac{b_i}{d_i},\frac{b_j}{d_j} \right) $, where $b_i, b_j$ and $d_i, d_j$ are the bandwidths and the degrees of nodes $i$ and $j$, respectively. Fig~\ref{fig:consensus_error_heter} shows how the consensus errors on different topologies evolve over time in the heterogeneous bandwidth scenario, where $b$ represents the ratio of the minimum available bandwidth on edges in the topology to the unit bandwidth of 3.25GB/s. It can be seen that when the number of edges is 32 and 48, BA-Topo outperforms existing network topologies. Moreover, when the number of edges is 16, the consensus speed of BA-Topo is slightly slower than exponential topology which has 32 edges, but still superior to other existing network topologies.

\subsubsection{Intra-Server Link Bandwidth Heterogeneity} \label{Intra-Server Link Bandwidth Heterogeneity}
\begin{figure}[!htpb]
    \centering
    \includegraphics[width=0.9\linewidth]{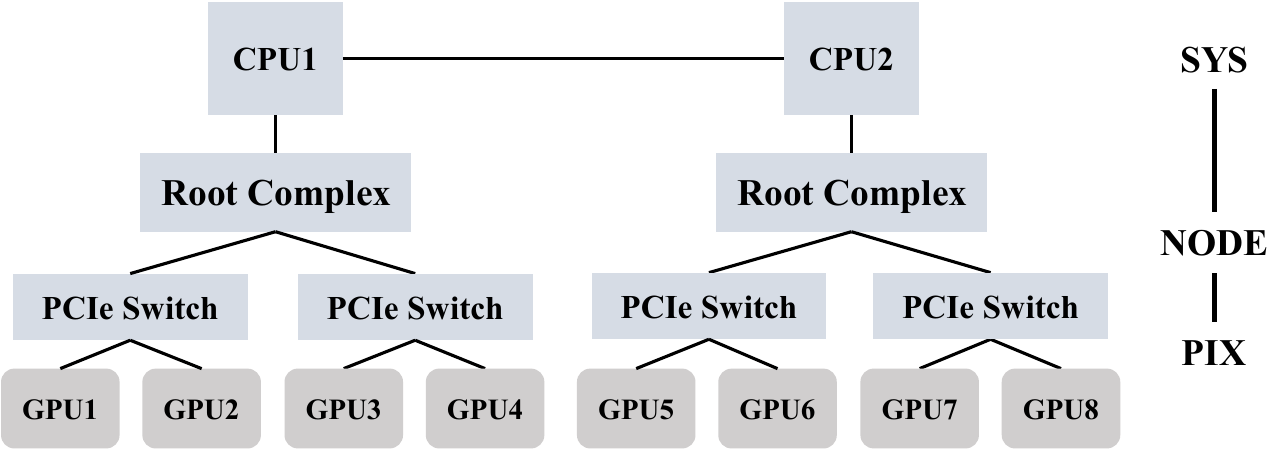}
    \caption{\label{Standard server architecture}Standard server architecture}
\end{figure}

We consider the standard server architecture as illustrated in Fig.~\ref{Standard server architecture}, 
where multiple types of intra-server physical interconnects exhibit heterogeneous available bandwidths. 
Note that, due to the hardware specifications of servers, each physical link can support only a limited number of concurrent logical edges.

Given the above standard server architecture, the edge-capacity constraint vector in Eq.~\eqref{tab:heterogeneous-problem-2} can be specified as
$\boldsymbol{e}=
\left(
e_{\mathrm{PIX1}},\,
e_{\mathrm{PIX2}},\,
e_{\mathrm{PIX3}},\,
e_{\mathrm{PIX4}},\,
e_{\mathrm{NODE1}},\,
e_{\mathrm{NODE2}},\,
e_{\mathrm{SYS}}
\right)^{T}$,
where each entry denotes the maximum number of logical edges that can be assigned to the corresponding interconnect. In this work, these limits evaluate to $\boldsymbol{e}=\left(1,\,1,\,1,\,1,\,4,\,4,\,16\right)^{T}.$
% Similarly, the  incidence matrix $M$ defined \yp{in Eq.~\eqref{tab:M_definition}} is instantiated in the same order as $\boldsymbol{e}$ by stacking the binary incidence vectors associated with each physical interconnect:
% \begin{equation}
% \label{tab:equ-3-26}
% M =
% \begin{bmatrix}
%     \boldsymbol{m}_{\mathrm{PIX1}}\\
%     \boldsymbol{m}_{\mathrm{PIX2}}\\
%     \boldsymbol{m}_{\mathrm{PIX3}}\\
%     \boldsymbol{m}_{\mathrm{PIX4}}\\
%     \boldsymbol{m}_{\mathrm{NODE1}}\\
%     \boldsymbol{m}_{\mathrm{NODE2}}\\
%     \boldsymbol{m}_{\mathrm{SYS}}
% \end{bmatrix}
% \in \{0,1\}^{7 \times |\mathcal{E}|},
% \end{equation}
% where each row vector $\boldsymbol{m}_{\ell} \in \{0,1\}^{|\mathcal{E}|}$, for $\ell\in\{\mathrm{PIX1},\mathrm{PIX2},\mathrm{PIX3},\mathrm{PIX4},\mathrm{NODE1},\mathrm{NODE2},\mathrm{SYS}\}$, indicates which logical edges are routed through the corresponding physical link.
The incidence matrix $M$ defined in Eq.~\eqref{tab:M_definition} is constructed by properly stacking the binary incidence vectors of the physical 
interconnects, where each row $\boldsymbol{m}_{\ell}\in\{0,1\}^{|\mathcal{E}|}$, 
$\ell\in\{\mathrm{PIX1},\mathrm{PIX2},\mathrm{PIX3},\mathrm{PIX4},\mathrm{NODE1},\mathrm{NODE2},\mathrm{SYS}\}$, 
indicates the logical edges occupying the physical link $\ell$.

\begin{figure}[!htpb]
  \centering
  \includegraphics[width=0.85\linewidth]{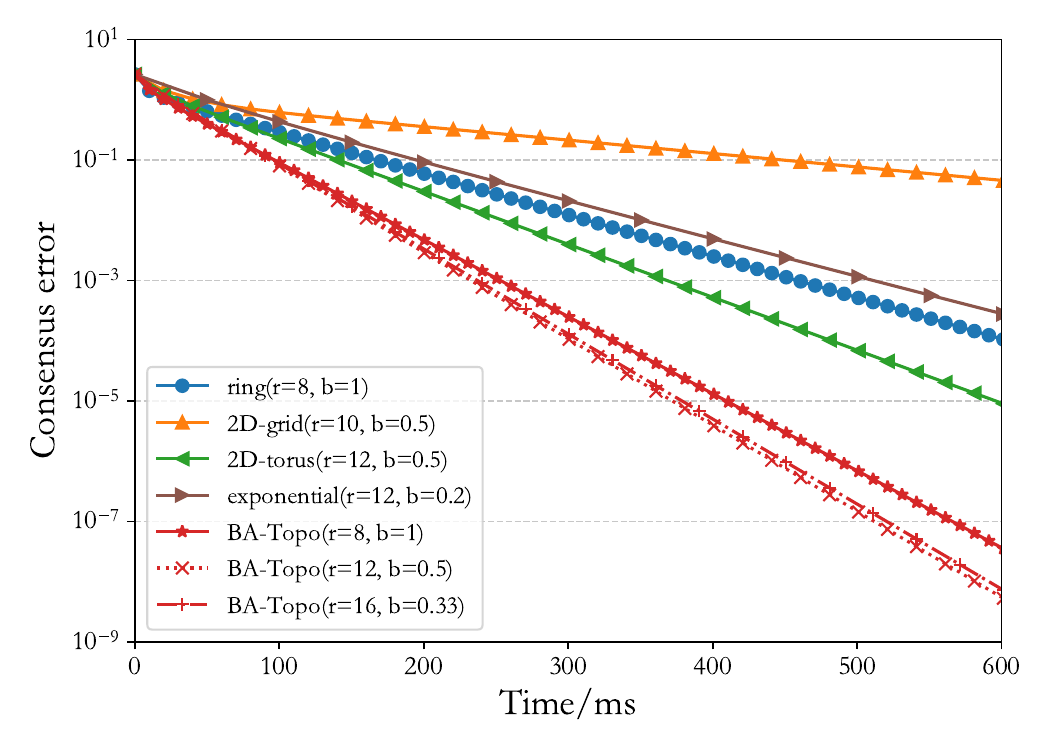}
  \caption{Comparison of consensus speed among various topologies with $n=8$ in intra-server link bandwidth heterogeneity scenario.}
  \label{fig:bandwidth-constraint-intra-consensus}
\end{figure}

We set the bandwidth ratio of the three types of physical links to be $b_{\mathrm{PIX}} : b_{\mathrm{NODE}} : b_{\mathrm{SYS}} = 1 : 1 : 2$, where the unit bandwidth corresponding to 1 is $4.88\ \mathrm{GB/s}$ and the bandwidth corresponding to 2 is $9.76\ \mathrm{GB/s}$. The available bandwidth of an edge $\{i,j\}$ is given by$\frac{b_{\mathrm{link}}}{e_{\mathrm{link}}}$,where $b_{\mathrm{link}}$ denotes the bandwidth of the physical link type to which edge $\{i,j\}$ belongs, and $e_{\mathrm{link}}$ denotes the number of edges mapped onto that physical link.
After time normalization for both the iteration time in the consensus experiments and the epoch time in the distributed deep learning experiments, the evolution of the consensus error under different parameter synchronization topologies is shown in Fig~\ref{fig:bandwidth-constraint-intra-consensus}. In the figure, $b$ denotes the ratio between the minimum available bandwidth among all edges in the topology and the unit bandwidth of $4.88\ \mathrm{GB/s}$.
It can be observed that, within the same time period, the consensus error of BA-Topo is smaller than that of existing parameter synchronization topologies. Moreover, in this heterogeneous-bandwidth setting, the performance of the exponential topology degrades, because the exponential topology maps 10 edges onto the SYS physical link, resulting in a minimum available edge bandwidth of only $0.976\ \mathrm{GB/s}$. This significantly increases the iteration time.

\subsubsection{Inter-Server Switch Port Bandwidth Heterogeneity} \label{Inter-Server Switch Port Bandwidth Heterogeneity}

\begin{figure}[!htpb]
    \centering
    \includegraphics[width=0.9\linewidth]{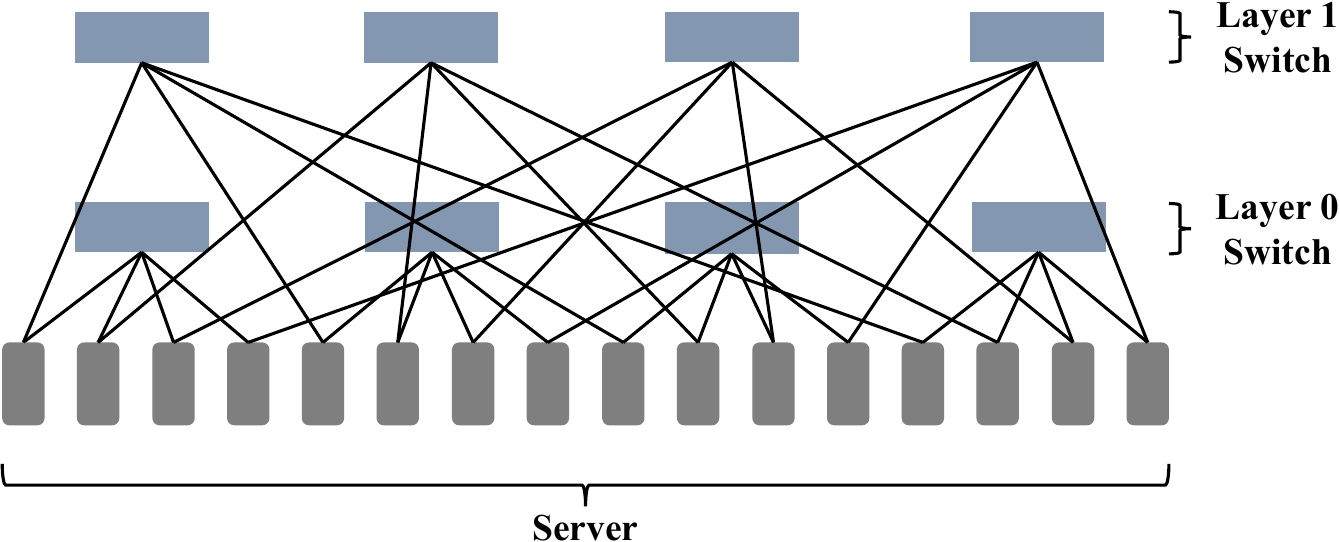}
    \caption{\label{BCube topology structure}BCube topology structure for $p=4$ and $k=2$.}
\end{figure}

For inter-server communication, we instantiate the general multi-layer 
switch-port model using the BCube architecture~\cite{guo2009bcube}. 
As illustrated in Fig.~\ref{BCube topology structure}, BCube is a server-centric 
topology where servers are interconnected through multiple layers of switches. 
In a BCube($p,k$) network, the total number of servers is $n=p^{k}$, each switch 
layer contains $p^{k-1}$ switches, and every server is equipped with $k$ network 
interfaces, each connecting to one switch at a distinct layer.
Note that switch ports at different BCube layers exhibit heterogeneous bandwidths. We 
denote the per-layer port bandwidths by $\boldsymbol{b} = ( b_{s_0}, \ldots, b_{s_{k-1}} )^{T}$.

Given that each switch has $p$ ports and each server attached to that switch can communicate with $p-1$ peer servers through it, the maximum number of logical edges that can be assigned to any switch port is $e_{s_i}=p-1,i=0,\ldots,k-1$.
Since each server is equipped with one port per layer and each layer contains $n=p^{k}$ ports in total, the edge-capacity constraint vector can be represented as $\boldsymbol{e}=\left(e_{s_0}\mathbf{1}_{n}^{T},\,e_{s_1}\mathbf{1}_{n}^{T},\,\ldots,\,e_{s_{k-1}}\mathbf{1}_{n}^{T}\right)^{T}$.

The general incidence matrix $M$ also specializes to the BCube topology. For each layer $i$, the block matrix $M_{s_i}$ defined in Eq.~\eqref{tab:heter-inter-M_s} encodes the BCube-specific server groupings: 
a row vector $\boldsymbol{m}_{s_ip_j}$ has the value $1$ at entries corresponding to logical edges whose endpoints are attached to the $j$-th port of the layer-$i$ switch, and $0$ elsewhere.
Stacking $\{M_{s_i}\}_{i=0}^{k-1}$ produces the full incidence matrix $M$. Substituting these BCube-specific $\boldsymbol{e}$ and $M$ into the general framework yields the optimization problem for the inter-server switch-port bandwidth heterogeneity experiments.

\begin{figure}[!htpb]
  \centering
  \includegraphics[width=0.85\linewidth]{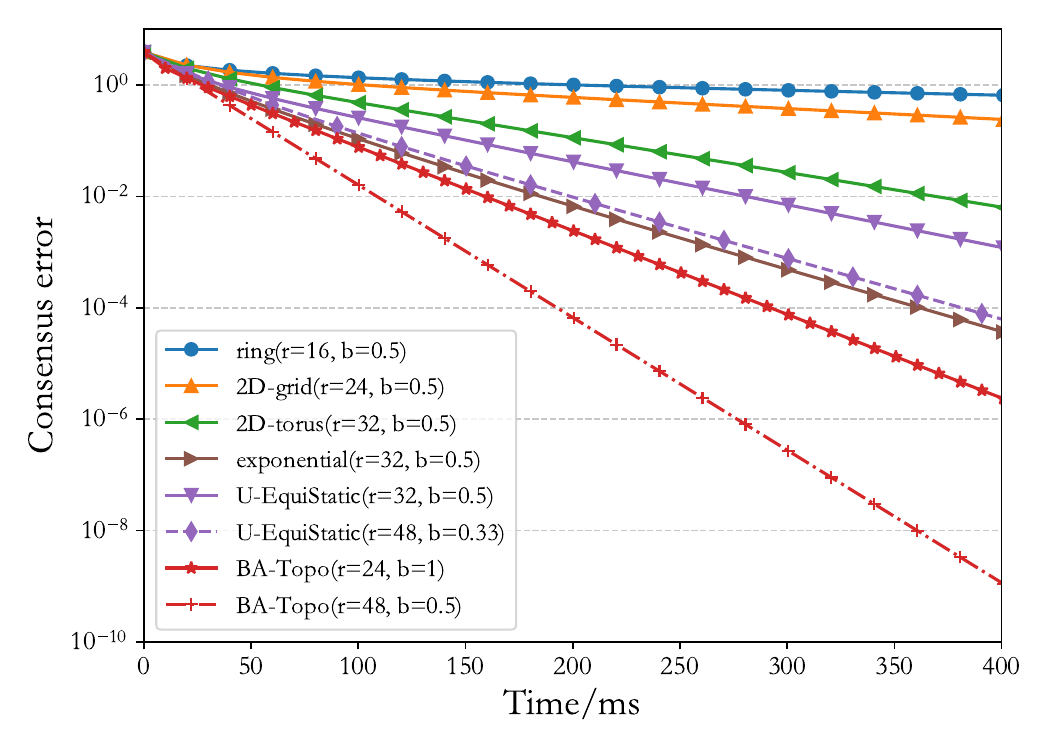}
  \caption{Comparison of consensus speed among various topologies with $n=16$ in inter-server switch port bandwidth heterogeneity scenario.}
  \label{fig:bandwidth-constraint-inter-consensus}
\end{figure}

We set the number of nodes to 16, the number of switch layers to 2, and the number of ports per switch to 4. The bandwidth ratios of switch ports at different layers are configured as $1:2$ and $2:3$, respectively.
Next, we conduct consensus experiments under the switch-port bandwidth ratio of $1:2$. Here, the bandwidth corresponding to 2 is $9.76\ \mathrm{GB/s}$, and the bandwidth corresponding to 1 is $4.88\ \mathrm{GB/s}$. The available bandwidth of an edge $\{i,j\}$ is given by
$\frac{b_{s_{ij}}}{e_{s_{ij}}}$,
where $b_{s_{ij}}$ denotes the bandwidth of the switch port traversed by edge $\{i,j\}$, and $e_{s_{ij}}$ denotes the number of edges mapped to that switch port.
After normalizing the iteration time of the consensus experiments, the evolution of the consensus error under different parameter synchronization topologies is shown in Fig~\ref{fig:bandwidth-constraint-inter-consensus}. In the figure, $b$ denotes the ratio between the minimum available bandwidth among all edges in the topology and the unit bandwidth of $4.88\ \mathrm{GB/s}$. It can be observed that, for the same elapsed time, BA-Topo achieves smaller consensus error compared with the other existing parameter synchronization topologies.

\subsubsection{Scalability} \label{Scalability}
% In the above experiments, exponential topology and U-EquiStatic topology have relatively fast consensus speeds among the existing topologies. 
To validate the scalability of our algorithm and compare the consensus speed, we evaluate the asymptotic convergence factor and the convergence time required for the consensus error to reach 0.0001 across different numbers of nodes in a homogeneous setting. We ensure that the degree sum of BA-Topo is always half of that of the exponential graph, and the degree sum of the U-EquiStatic topology is close to half of that of the exponential graph to maintain similar sparsity among the topologies. The experimental results, as shown in Tab.~\ref{tab:iso-bandwidth-consensus}, indicate that as the number of nodes increases from 4 to 128, the asymptotic convergence factor and convergence time of BA-Topo are generally smaller than those of the exponential topology and U-EquiStatic topology. This demonstrates that BA-Topo performs almost the best in terms of consensus speed at different scales.

\subsection{Decentralized Learning} \label{decentralized-experiments}
Next, we validate the effectiveness of BA-Topo in decentralized learning. We use DSGD~\cite{lian2017can} as the distributed training algorithm, with CIFAR-10 and CIFAR-100~\cite{krizhevsky2009learning} as the training datasets, and ResNet-18~\cite{he2016deep} as the training model. 
Before the training, each node randomly samples the same number of samples from each class of training data to form its local training dataset. 
As for the hyperparameters of the experiment, the batch size is set to 32 (per node), the learning rate is set to 0.05, the momentum factor is set to 0.9, the weight decay is set to 0.0001, and the training epoch is set to 100.

\begin{table*}[!t]
\centering
\caption{Comparison of DSDG over various topologies in terms of training time (seconds) under four bandwidth scenarios.}
\label{tab:all-scenarios-compact}
\setlength{\tabcolsep}{2mm}
\scalebox{0.85}{
\begin{tabular}{c|c|cccccc|cccc}
\toprule[2pt]
Dataset & Scenario 
& Ring & 2D-Grid & 2D-Torus & Exponential 
& \makecell[c]{U-EquiStatic\\($r$=32)} 
& \makecell[c]{U-EquiStatic\\($r$=48)}
& \makecell[c]{\textbf{BA-Topo}\\(1)} 
& \makecell[c]{\textbf{BA-Topo}\\(2)} 
& \makecell[c]{\textbf{BA-Topo}\\(3)} 
& \makecell[c]{\textbf{BA-Topo}\\(4)} \\
\midrule

% ================= CIFAR10 =================
\multirow{4}{*}{CIFAR-10}

& Homogeneous
& 229.7 & 227.7 & 189.8 & 134.6 
& 138.0 & 159.5 
& 165.4 (r=16) 
& 144.9 (r=24) 
& \textbf{110.1 (r=32)} 
& 146.1 (r=54) \\

& Node-Level
& 412.0 & 486.4 & 405.3 & 287.4 
& 294.8 & 371.2 
& 275.3 (r=16) 
& \textbf{181.6 (r=32)} 
& 182.9 (r=48) 
& -- \\

& Intra-Server
& 484.1 & 662.4 & 407.1 & 754.0 
& -- & -- 
& \textbf{266.8 (r=8)}
& 276.0 (r=12) 
& 407.6 (r=16) 
& -- \\

& Inter-Server
& 320.9 & 227.7 & 189.8 & 134.6 
& 138.0 & 159.5 
& 101.3 (r=24) 
& \textbf{86.3 (r=48)} 
& -- 
& -- \\
\midrule

% ================= CIFAR100 =================
\multirow{4}{*}{CIFAR-100}

& Homogeneous
& 234.7 & 269.1 & 200.1 & 151.8 
& 141.5 & 190.5 
& 177.8 (r=16) 
& 172.5 (r=24) 
& \textbf{127.6 (r=32)} 
& 183.9 (r=54) \\

& Node-Level
& 420.9 & 574.9 & 427.5 & 324.3 
& 302.2 & 443.3 
& 239.76 (r=16) 
& 248.1 (r=32) 
& \textbf{194.7 (r=48)} 
& -- \\

& Intra-Server
& 350.7 & 552.0 & 510.6 & 651.8 
& -- & -- 
& \textbf{261.8 (r=8)}
& 269.1 (r=12) 
& 310.1 (r=16) 
& -- \\

& Inter-Server
& 327.8 & 269.1 & 200.1 & 151.8 
& 141.5 & 190.5 
& 126.0 (r=24) 
& \textbf{117.3 (r=48)} 
& -- 
& -- \\
\bottomrule[2pt]
\end{tabular}}
\\[1mm]
\begin{minipage}{\textwidth}
\footnotesize
\raggedright
\yp{
$^*$For each bandwidth scenario, multiple BA-Topo configurations with 
different total numbers of edges are evaluated. The feasible number of edges varies 
across scenarios due to the distinct physical bandwidth constraints imposed by nodes, 
intra-server links, and inter-server switch ports.
}
\end{minipage}
\end{table*}

We also evaluate the time required for each epoch of training by simulation. First, we test the computation time per iteration for training ResNet-18 on a single 2080Ti, which is 15.21ms (denoted as $t_{comp}$), and the communication time for exchanging parameters of ResNet-18 over a link with a bandwidth of 9.76GB/s, which is 5.01ms (denoted as $t_{comm}$). Then, the smallest available bandwidth in all edges is calculated to scale the training time of one epoch:
\begin{equation}
        \label{tab:epoch-time-scale}
        % t_{epoch}=\left( \frac{9.76}{b_{min}}\times 5.01+15.21 \right) \times c_{iter},
        t_{epoch}=\left( \frac{b_{avail}}{b_{min}}\times t_{comm}+t_{comp} \right) \times c_{iter},
\end{equation}
where $b_{min}$ represents the minimal bandwidth of all edges and $c_{iter}$ represents the number of iterations in one epoch.

\subsubsection{Homogeneous Bandwidth}
\begin{figure*}[!t]
    \centering
    \subfigure[CIFAR-10]{
        \includegraphics[width=0.4\linewidth]{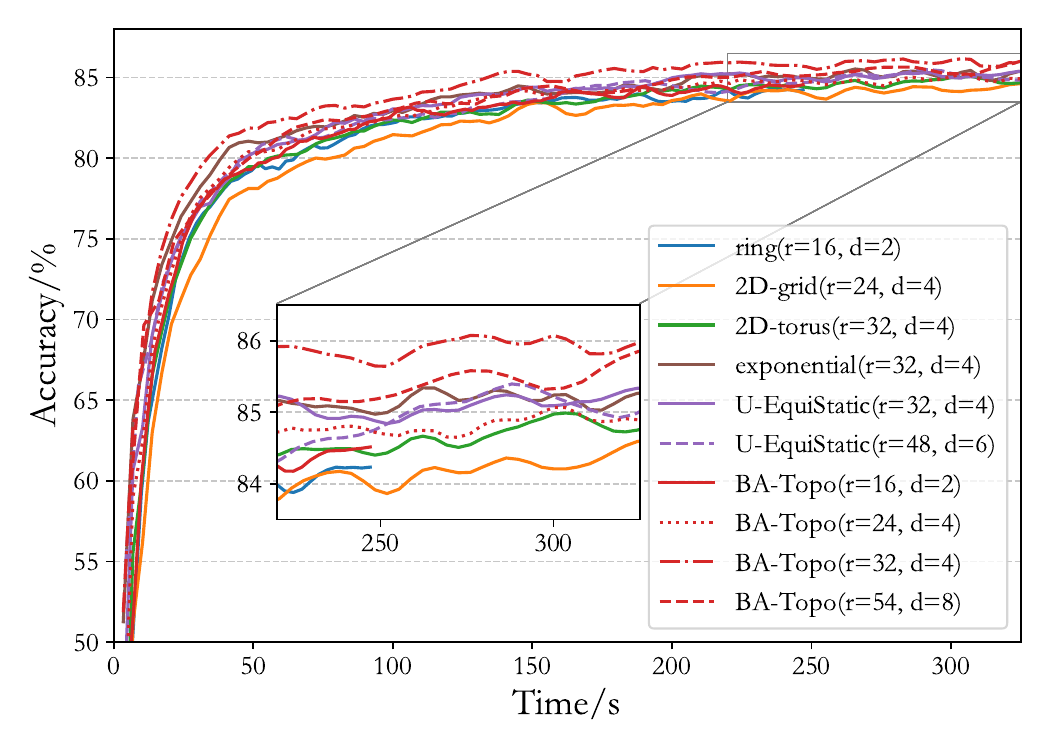}
    }
    \subfigure[CIFAR-100]{
        \includegraphics[width=0.4\linewidth]{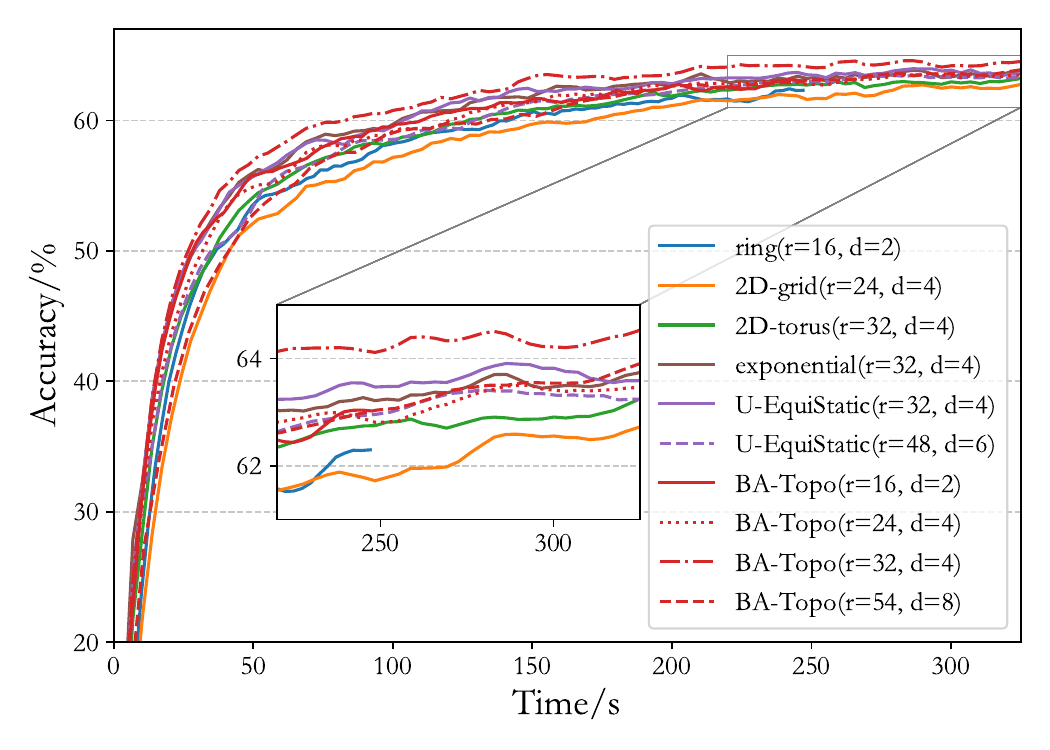}
    }
    \caption{\label{fig:Accuracy-homo}Test accuracy of DSGD on various topologies with $n=16$ in homogeneous bandwidth scenario.}
\end{figure*}

The number of nodes and bandwidth settings are the same as described in Sec.~\ref{homo-consensus}. We compare the test accuracy of DSGD on different topologies and the experimental results are shown in Fig.~\ref{fig:Accuracy-homo}. It can be seen that when the number of edges is the same, the prediction accuracy on BA-Topo is higher than existing network topologies within the same time frame. Moreover, among all topologies, BA-Topo achieves the highest prediction accuracy when the number of edges is 32. The curves corresponding to the ring topology and BA-Topo ($r$=16, $d$=2) disappear around 250s, indicating that these two topologies have completed training for 100 epochs.

\begin{figure*}[!t]
    \centering
    \subfigure[CIFAR-10]{
        \includegraphics[width=0.4\linewidth]{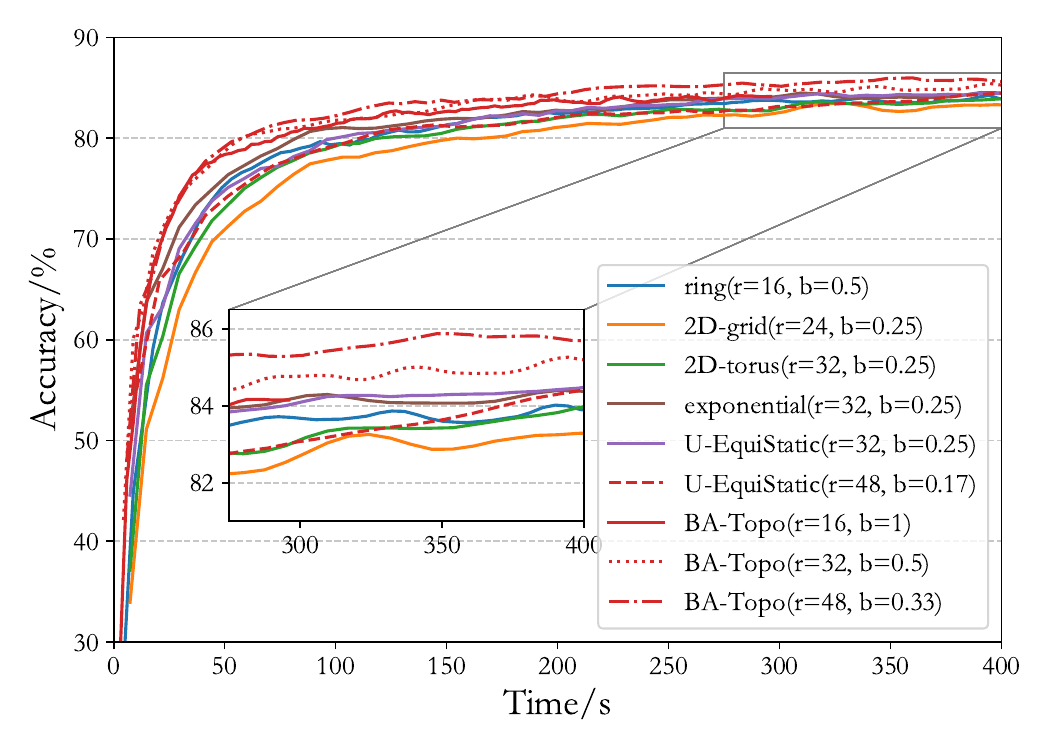}
    }
    \subfigure[CIFAR-100]{
        \includegraphics[width=0.4\linewidth]{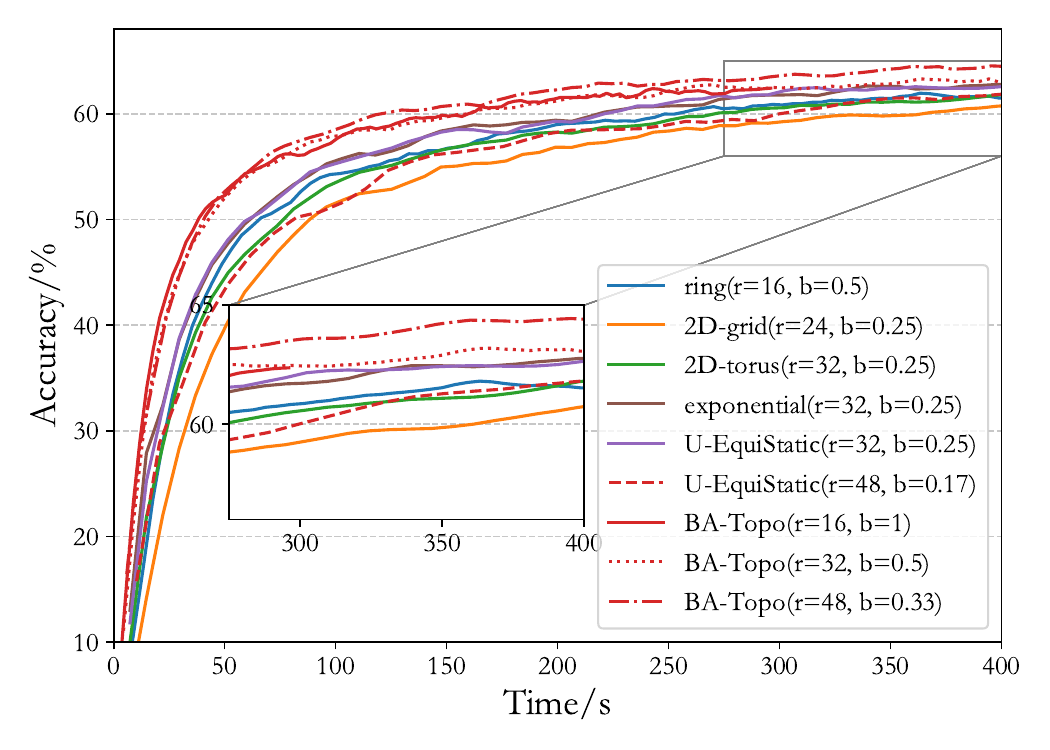}
    }
    \caption{\label{fig:Accuracy-heter-node}Test accuracy of DSGD on various topologies with $n=16$ in node-level bandwidth heterogeneity scenario.}
\end{figure*}

\begin{figure*}[!t]
    \centering
    \subfigure[CIFAR-10]{
        \includegraphics[width=0.4\linewidth]{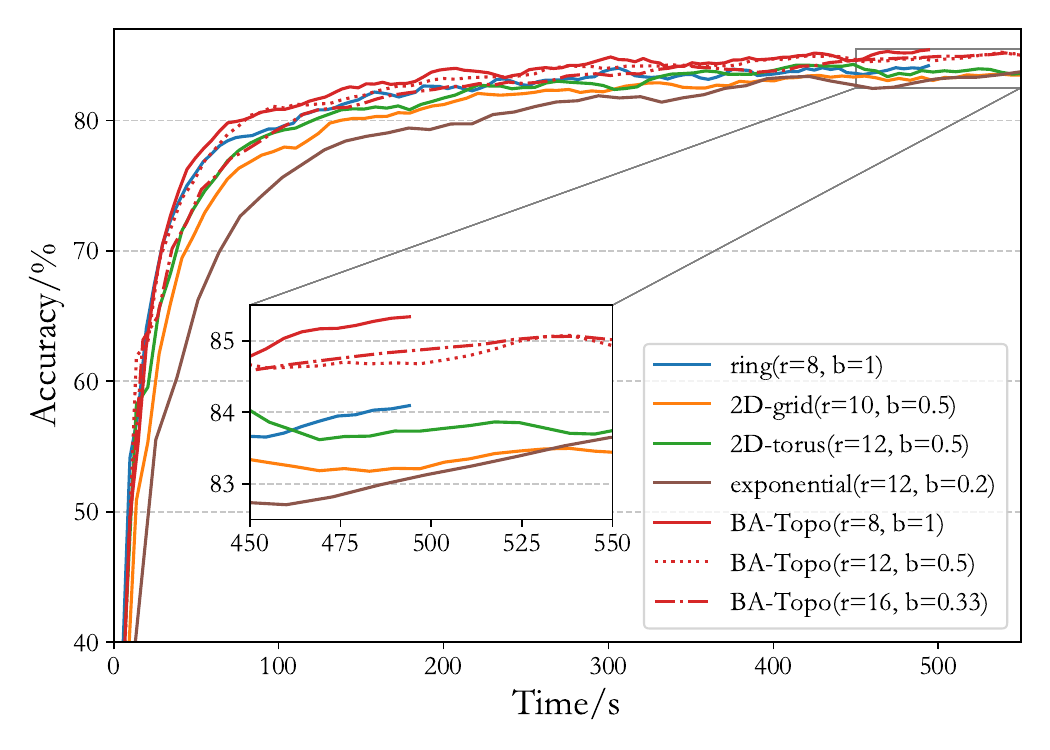}
    }
    \subfigure[CIFAR-100]{
        \includegraphics[width=0.4\linewidth]{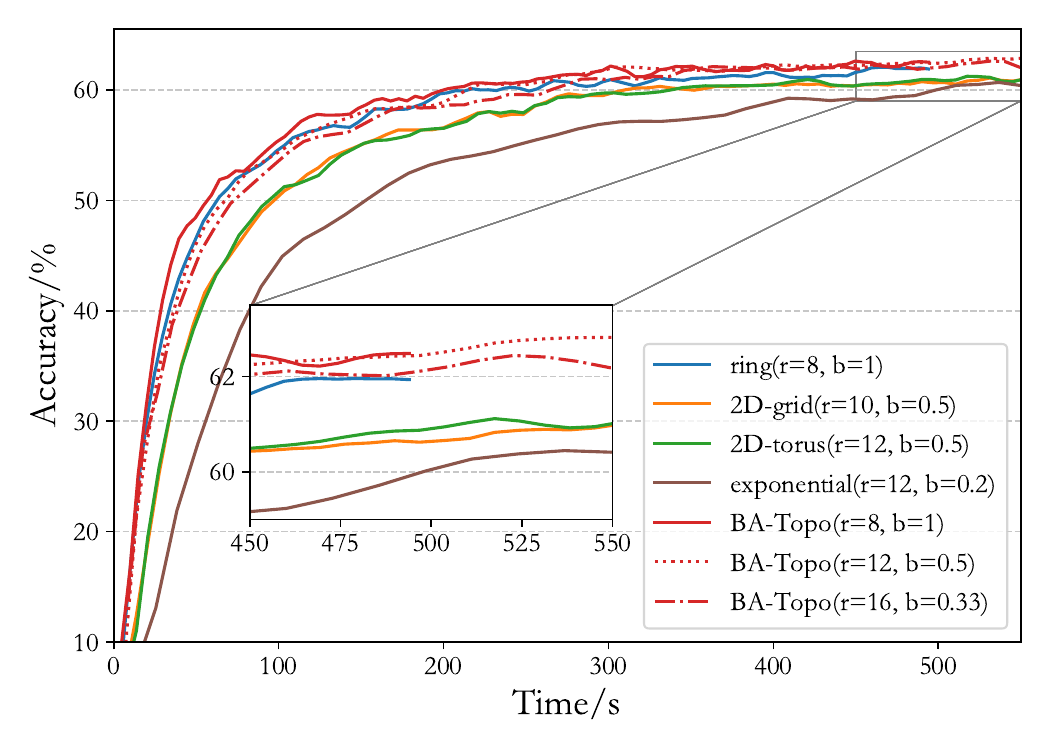}
    }
    \caption{\label{fig:Accuracy-heter-intra}Test accuracy of DSGD on various topologies with $n=8$ in intra-server link bandwidth heterogeneity scenario.}
\end{figure*}

\begin{figure*}[!t]
    \centering
    \subfigure[CIFAR-10]{
        \includegraphics[width=0.4\linewidth]{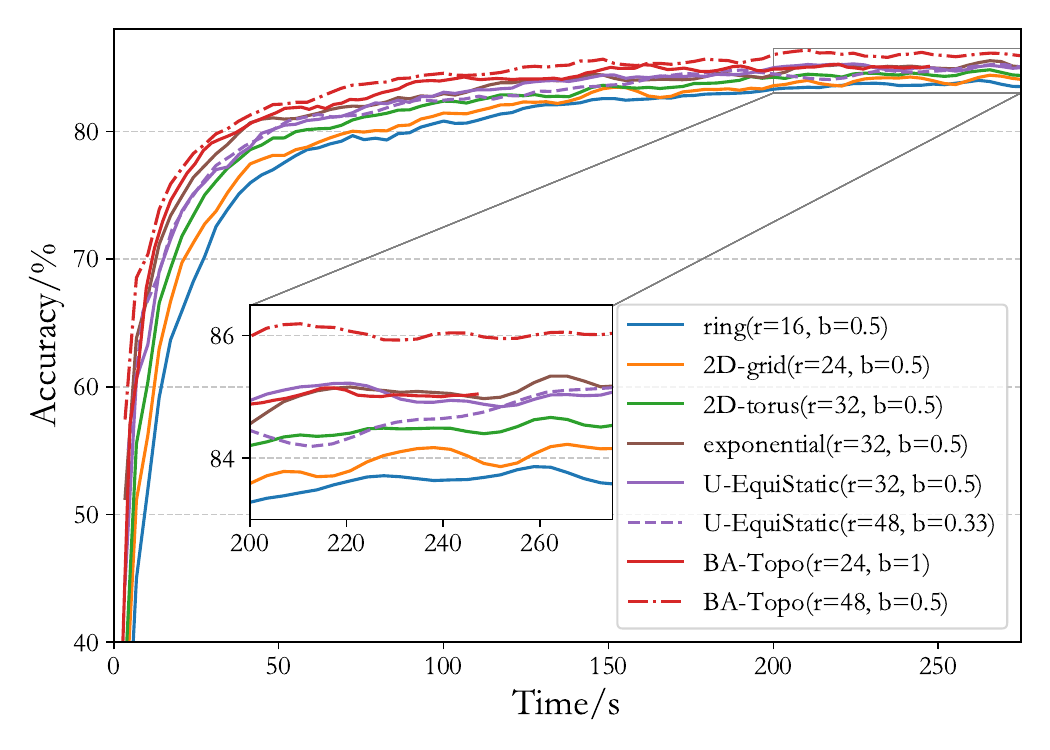}
    }
    \subfigure[CIFAR-100]{
        \includegraphics[width=0.4\linewidth]{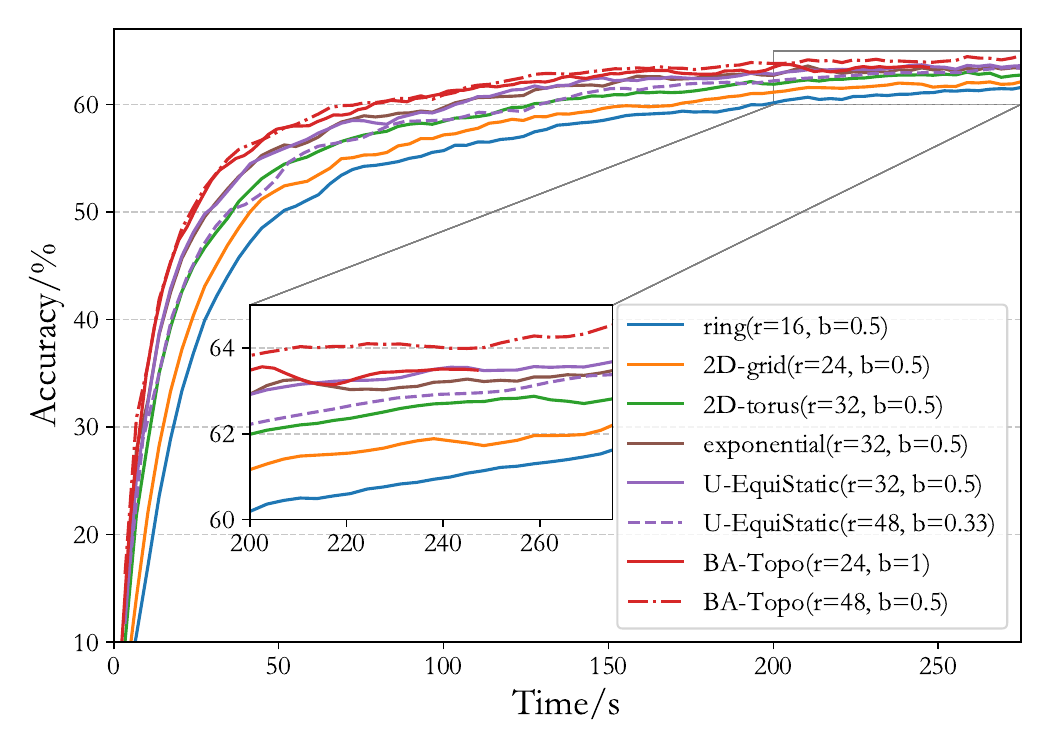}
    }
    \caption{\label{fig:Accuracy-heter-inter}Test accuracy of DSGD on various topologies with $n=16$ in inter-server switch port bandwidth heterogeneity scenario.}
\end{figure*}

In order to quantitatively evaluate the impact of different network topologies on the training performance of decentralized learning, we test the time required for DSGD to reach a testing accuracy of 84\% on CIFAR-10 and 62\% on CIFAR-100\footnote{Under the experimental settings in this section, 84\% and 62\% are relatively high prediction accuracy values achievable by all topologies.}, as shown in Tab.~\ref{tab:all-scenarios-compact}. 
\yp{
It follows that when the number of edges is 32, BA-Topo achieves the fastest convergence among all topologies: it reaches the target test accuracy in 110.1s on CIFAR-10 and 127.6s on CIFAR-100. These results correspond to speedups of at least $1.22\times$ and $1.11\times$, respectively, over the other topologies.
% at least 9.8\% faster than other topologies. 
}
The reason why BA-Topo with other numbers of edges do not perform as well as BA-Topo ($r$ = 32) is that when there are fewer edges, the topology becomes sparser, resulting in slower consensus speed. On the other hand, with more edges, nodes have higher degrees, leading to increased communication time per iteration.

\subsubsection{Node-Level Bandwidth Heterogeneity}
The number of nodes and bandwidth settings are the same as described in Sec.~\ref{Node-Level Bandwidth Heterogeneity}. The experimental results are shown in Fig.~\ref{fig:Accuracy-heter-node}. It can be seen that within the same time frame, the prediction accuracy on BA-Topo with different numbers of edges is higher than that on existing network topologies. Although the consensus speed of BA-Topo ($r$=16, $b$=1) is slower than the exponential topology and the U-EquiStatic topology, the higher available bandwidth on edges in this topology results in shorter time per iteration in decentralized learning, which allows for more model parameter update steps within the same time frame, accelerating the convergence speed of the model.

Moreover, we test the time required for DSGD to reach a testing accuracy of 84\% on CIFAR-10 and 62\% on CIFAR-100, and experimental results are shown in Tab.~\ref{tab:all-scenarios-compact}. It can be seen that the time required for the test accuracy to reach the set value on BA-Topo with different numbers of edges is significantly less than that of existing network topologies. 
% on CIFAR-10 and CIFAR-100, BA-Topo ($r$ = 32) is at least 36.8\% and 17.9\% faster than existing topologies, and BA-Topo ($r$ = 48) is at least 36.4\% and 35.6\% faster than existing topologies, 
\yp{
Specifically, 
for CIFAR-10, BA-Topo ($r$ = 32) achieves a speedup of at least $1.58\times$ over the other topologies, 
while BA-Topo ($r$ = 48) achieves a speedup of at least $1.55\times$ over other topologies for CIFAR-100. 
% In conclusion, BA-Topo ($r$ = 48) performs best among all the topologies.
}

\subsubsection{Intra-Server Link Bandwidth Heterogeneity}
The number of nodes and bandwidth settings are the same as described in Sec.~\ref{Intra-Server Link Bandwidth Heterogeneity}. The experimental results are shown in Fig.~\ref{fig:Accuracy-heter-intra}. It can be seen that within the same time frame, the prediction accuracy on BA-Topo with different numbers of edges is higher than that on existing network topologies. 

Moreover, we test the time required for DSGD to reach a testing accuracy of 84\% on CIFAR-10 and 61\% on CIFAR-100. \yp{It follows from Tab.~\ref{tab:all-scenarios-compact} that BA-Topo ($r$ = 8) achieves speedups of at least $1.53\times$ and $1.34\times$ over the other topologies for CIFAR-10 and CIFAR-100, respectively.
}

\subsubsection{Inter-Server Switch Port Bandwidth Heterogeneity}
The number of nodes and bandwidth settings are the same as described in Sec.~\ref{Inter-Server Switch Port Bandwidth Heterogeneity}. The experimental results are shown in Fig.~\ref{fig:Accuracy-heter-inter}. It can be seen that within the same time frame, the prediction accuracy on BA-Topo with different numbers of edges is higher than that on existing network topologies. 

Moreover, we test the time required for DSGD to reach a testing accuracy of 84\% on CIFAR-10 and 62\% on CIFAR-100. 
\yp{It follows from Tab.~\ref{tab:all-scenarios-compact} that BA-Topo ($r$ = 48) achieves speedups of at least $1.56\times$ and $1.21\times$ over the other topologies for CIFAR-10 and CIFAR-100 respectively.
}

\section{Conclusions} \label{section7}
In this paper, we have proposed a bandwidth-aware network topology optimization framework for decentralized learning. The proposed framework differs from existing works by formulating the design of the network topology as an optimization problem with cardinality constraints on the edge set, enabling it to handle both homogeneous and heterogeneous bandwidth scenarios. To enhance solvability, we reformulate the problem into an equivalent Mixed-Integer SDP problem which can be efficiently solved under an ADMM-based framework, yielding a superior network topology that accelerates consensus and speeds up decentralized learning on real-world datasets compared to the benchmark topologies.
Future work will focus on addressing dynamic bandwidth scenarios with a time-varying network topology optimization solution.

\bibliographystyle{IEEEtran}
\bibliography{IEEEabrv,ref}
\end{document}